\documentclass[twocolumn,preprintnumbers,superscriptaddress,nofootinbib,prd]{revtex4-1}
\usepackage{graphicx,latexsym,amsfonts,amssymb,amsmath,slashed,array,dcolumn,feynmp}

\pdfoutput=1

\def\lsim{\mathrel{\raise.3ex\hbox{$<$\kern-.75em\lower1ex\hbox{$\sim$}}}}
\def\gsim{\mathrel{\raise.3ex\hbox{$>$\kern-.75em\lower1ex\hbox{$\sim$}}}}

\usepackage{xcolor}
\definecolor{red}{rgb}{1.0, 0, 0}

\newcommand{ \slashchar }[1]{\setbox0=\hbox{$#1$}   % set a box for #1
   \dimen0=\wd0                                     % and get its size
   \setbox1=\hbox{/} \dimen1=\wd1                   % get size of /
   \ifdim\dimen0>\dimen1                            % #1 is bigger
      \rlap{\hbox to \dimen0{\hfil/\hfil}}          % so center / in box
      #1                                            % and print #1
   \else                                            % / is bigger
      \rlap{\hbox to \dimen1{\hfil$#1$\hfil}}       % so center #1
      /                                             % and print /
   \fi}                                             %

\newcommand{\gev}{\text{GeV}}

\newcommand{\fb}{\text{fb}}

\newcommand{\ra}{\rightarrow}
\newcommand{\AP}{\mathcal{P}}
\newcommand{\AQ}{\mathcal{Q}}
\newcommand{\mh}{m_h}

\begin{document}

\title{Enhanced di-Higgs Production through Light Colored Scalars}

\author{Graham D. Kribs}
\affiliation{Department of Physics, University of Oregon,
             Eugene, OR 97403}

\author{Adam Martin}
\affiliation{Theoretical Physics Department, Fermilab, Batavia, IL 60510}
\affiliation{Department of Physics, University of Notre Dame, Notre Dame, IN 46556\,}
~\email{visiting scholar}

\preprint{FERMILAB-PUB-12-395-T}
\date{\today}

\begin{abstract}

We demonstrate enhanced di-Higgs production at the LHC in the presence
of modifications of the effective couplings of Higgs to gluons 
from new, light, colored scalars.  While our results apply 
to an arbitrary set of colored scalars, 
we illustrate the effects with a real color octet scalar -- 
a simple, experimentally viable model involving a 
light ($\simeq 125$-$300$~GeV) colored scalar.
Given the recent LHC results, we consider two distinct scenarios:
First, if the Higgs is indeed near $125$~GeV, we show that the 
di-Higgs cross section could be up to nearly $10^3$ times the
Standard Model rate for particular octet couplings and masses.
This is potentially observable in \emph{single} Higgs production modes, 
such as $pp \ra h h \ra \gamma\gamma b\bar{b}$ as well as
$pp \ra h h \ra \tau^+\tau^- b\bar{b}$ where a small fraction 
of the $\gamma\gamma$ or $\tau^+\tau^-$ events near the 
putative Higgs invariant mass peak contain also a $b\bar{b}$ 
resonance consistent with the Higgs mass.
Second, if the Higgs is not at $125$~GeV (and what the LHC has observed
is an impostor), we show that the same parameter region 
where singly-produced Higgs production can be suppressed below 
current LHC limits, for a heavier Higgs mass, also simultaneously 
predicts substantially enhanced di-Higgs production.  
We point out several characteristic signals of di-Higgs production
with a heavier Higgs boson, such as $pp \ra hh \ra W^+W^-W^+W^-$,
which could use same-sign dileptons or trileptons plus missing energy 
to uncover evidence.

\end{abstract}

\maketitle

%------------------------------------------------------------------------------
\section{Introduction}
\label{sec:intro}

Di-Higgs production in the Standard Model (SM) has a very small 
rate at the LHC 
\cite{Glover:1987nx,Plehn:1996wb,Dawson:1998py,Djouadi:1999rca}.
For $\mh = 125$~GeV,
the leading order cross section is 
$\sigma(p p \ra h h) = 4 \, (16)$~fb at $\sqrt{s} = 8 \, (14)$~TeV\@.
This is much smaller than the single Higgs production cross section
due to the larger partonic energy needed to produce two Higgs bosons,
as well as an accidental cancellation between
the $s$-channel and box diagram contributions to the 
amplitude $gg \ra hh$.
Hence, like other accidentally suppressed processes in the 
Standard Model, di-Higgs production provides a great opportunity 
for new physics to be observed. 
In this paper we demonstrate that in the presence of a 
a general set of light colored scalars, the di-Higgs production 
cross section can be
enhanced by orders of magnitude above the Standard Model rate.  
Light colored scalars that couple to the Higgs boson are 
well known to have a profound effect on 
\emph{single} Higgs production~\cite{susysuppression,Manohar:2006ga,Arnesen:2008fb,Ma:2011kc,arXiv:0709.4227,Boughezal:2010ry,Bai:2011aa,Dobrescu:2011aa,Batell:2011pz}.  
Di-Higgs production has been considered previously in several
different contexts \cite{Barger:1988kb,Hagiwara:1989xx,Barger:1991jn,Djouadi:1999gv,Belyaev:1999mx,BarrientosBendezu:2001di,Baur:2002rb,Baur:2002qd,Baur:2003gp,Moretti:2004wa,Plehn:2005nk,Binoth:2006ym,Pierce:2006dh,Dawson:2006dm,Moretti:2007ca,Kanemura:2008ub,Lafaye:2009vr,Arhrib:2009hc,Moretti:2010kc,Asakawa:2010xj,Grober:2010yv,Contino:2012xk,Dolan:2012rv}.

Since direct searches at the LHC for light colored scalars that couple 
to the Higgs have many constraints, depending on the model and the
decay modes, the burden is on us to provide a ``benchmark model'' that
evades these constraints while providing the enhancements in di-Higgs
production that we find. 
Several possible representations could be considered, including the
superpartners to the top, vector-like sets of particles 
(that may or may not carry electroweak quantum numbers), 
or real representations such as an electroweak neutral color octet.

A real scalar octet is an interesting example of a colored
scalar that can be effectively hidden in the LHC data.
In the model that we consider, these scalars are pair-produced 
(up to small corrections coming from loop processes), 
and decay through loop-level processes 
to a pair of gluons. Thus, the signature is two pairs of jets 
with equal invariant mass, making four total jets.  
ATLAS~\cite{Aad:2011yh} and CMS~\cite{CMSoctets} have multi-jet 
searches that are potentially sensitive to this scalar;
the ATLAS study restricts $M_S \gsim 125$~GeV, 
while the CMS study \cite{CMSoctets} is sensitive only for 
scalars exceeding $320$~GeV\@. 
This leaves a wide range of 
real scalar color octets are allowed by LHC constraints.  
A real scalar color octet $S_a$ can couple to the Higgs boson 
through the renormalizable Higgs-portal interaction of the form 
$(\kappa/2)S_a^2 H^\dagger H$.  Vital to our study is the
possibility that $\kappa$ can be negative, as we will see.

The correlation between effects in single-Higgs and di-Higgs 
production has been studied previously in terms of 
contact interactions~\cite{Pierce:2006dh,Kanemura:2008ub}. 
For very heavy colored scalars -- meaning scalars whose mass is large compared 
to the electroweak scale, $M_i \gg v$, as well as to the 
characteristic energy of di-Higgs production 
(roughly $M_i \gg \hat{s} \sim \mathcal{O}(10\, \mh^2)$ -- the 
effective operators provide a reasonable description.
As we will see, for the colored scalars we consider in this paper,
between about $100$ to $300$~GeV, the full momentum-dependence
of the one-loop calculation is essential to accurately 
estimate the di-Higgs enhancement.  We show this explicitly
in Appendix~\ref{sec:operatorcompare}.

Finally, it may seem somewhat quixotic that we consider
$\mh = 125$~GeV as well as $\mh > 125$~GeV, given the strong
evidence from the LHC for a new particle with properties 
consistent with a Higgs near $\mh = 125$~GeV \cite{atlashiggs,cmshiggs}.
While much work remains to be done to verify that the
$125$~GeV particle is indeed the (or a) Higgs boson,
there is a more direct reason for our continued interest
in a heavier Higgs boson.  As we showed in Ref.~\cite{Dobrescu:2011aa},
colored scalars can \emph{suppress} single Higgs production
well below the current bounds from LHC for a wide range of Higgs masses.
So the argument that ``we have already searched for a heavier Higgs boson
and did not find it, so the Higgs must be the $125$~GeV 
Higgs-like particle'' is simply wrong.  
This reasoning is wrong because the argument applies \emph{only} 
to the Standard Model.  The general class of models we consider in this
paper -- colored scalars with Higgs portal couplings -- provide 
a clear class of counterexamples.  Of course we are not disputing 
the strong evidence for a $125$~GeV particle; instead, we believe 
maintaining a healthy dose of skepticism regarding the true identity
of this particle, given the wide number of impostors~\cite{Higgsimpostors} 
that could be masquerading as a Higgs-like resonance.

%------------------------------------------------------------------------------
\section{Effective Couplings}
\label{sec:let}

Low energy theorems for Higgs physics provide powerful methods  
to determine the effective Higgs couplings to both Standard Model 
particles as well as new physics
\cite{Ellis:1975ap,Shifman:1979eb,Hagiwara:1989xx,Kniehl:1995tn,Low:2009di,Gillioz:2012se}.
We are interested in extending the Standard Model to include a general set of 
colored scalars $\phi_i$ in arbitrary representations of QCD.  
The multiplicity of scalars (number of flavors) is taken into account,
while our predictions for the single and di-Higgs rates are otherwise 
independent of the electroweak quantum numbers at leading order in the
couplings.  

The minimal Lagrangian for colored scalars in complex representations is
\begin{eqnarray}
\mathcal{L}_c &=&
(D_\mu \phi_i)^\dagger (D^\mu \phi_i) 
- m_i^2 \phi_i^\dagger\phi_i 
- \kappa_i \phi_i^\dagger\phi_i H^\dagger H
\end{eqnarray}
while for real scalars the Lagrangian is
\begin{eqnarray}
\mathcal{L}_r &=& 
\frac{1}{2} (D_\mu \phi_i)^2 
- \frac{1}{2} m_i^2 \phi_i^2 
- \frac{\kappa_i}{2} \phi_i^2 H^\dagger H \, .
\end{eqnarray}
We have not included quartic (or possibly cubic) self-interactions,
nor interactions among different \emph{flavors} of scalars,
since these couplings will not play any role in our calculations
of di-Higgs production.  The field-independent scalar mass-squared is
\begin{eqnarray}
M_i^2 & \equiv & m_i^2 + \frac{\kappa_i v^2}{2} \, ,
\end{eqnarray}
while the Higgs field-dependent mass is
\begin{eqnarray}
M_i^2(h) &=& m_i^2 + \frac{\kappa_i}{2} (v + h)^2 \, ,
\end{eqnarray}
both of which apply to scalars in real or complex representations.
When $\mh \ll M_i$, the colored scalars can be integrated out, resulting 
in an effective theory in powers of $1/M_i$.  The leading interactions
of the Higgs to gluons can be determined by matching the strong 
coupling constant in the low energy and high energy theory \cite{Low:2009di},
\begin{eqnarray}
{\cal L}_{\rm eff} 
   &=& -\frac{1}{4} \frac{1}{g^2_{\rm eff}(\mu, h)}
     G_{\rm \mu\nu}^a G^{a \, \mu\nu} \nonumber \\
   &=& - \frac{1}{4} \bigg[ 
         \frac{1}{g^2(\mu)} 
         - \frac{1}{12 \pi^2} \log \frac{m_t(h)}{\mu} \nonumber \\
   & &   \qquad - \sum_i \frac{C_i}{24 \pi^2} \log \frac{M_i(h)}{\mu} \bigg] 
         \, G_{\rm \mu\nu}^a G^{a \, \mu\nu} 
         \label{eq:lowratt}
\end{eqnarray}
where we have included the top quark and an arbitrary set of colored scalars
with explicit Higgs field-dependence.  
For a real scalar field, $C_i$ is the Dynkin index, 
e.g.~$C_i = 3$ for a color octet;
for a complex representation, replace $C_i \ra 2 C_i$.
This leads to the one-loop effective Lagrangian,
\begin{equation}
% \frac{g_s^2}{48 \pi^2} 
\frac{\alpha_s}{12 \pi} 
\left[ \log m_t(h) + \sum_i \frac{C_i}{2} \log M_i(h) \right] 
G_{\rm \mu\nu}^a G^{a \, \mu\nu} \, ,
\label{eq:eff}
\end{equation}
where we have shifted $G_{\mu\nu}^a$ back to the canonical basis.

Expanding Eq.~(\ref{eq:eff}) to first order in $h/v$, 
we obtain the single Higgs effective interaction with 
colored scalars, 
\begin{eqnarray}
\frac{h}{v} \, 
G_{\rm \mu\nu}^a G^{a \, \mu\nu} \, 
% \sum_i \frac{g_s^2}{192 \pi^2} C_i 
\sum_i \frac{\alpha_s}{48 \pi} C_i 
\frac{\kappa_i v^2}{M_i^2} \, . \label{eq:oneHiggseff}
\end{eqnarray}
The single Higgs effective interaction, 
Eq.~(\ref{eq:oneHiggseff}), is for example identical to the
result obtained from our previous study of Higgs suppression 
through colored scalars in the limit $\mh \ll M_i$ 
[Eq.~(2.7) from Ref.~\cite{Dobrescu:2011aa}].

The 4-point di-Higgs effective interaction is similarly obtained, 
\begin{eqnarray}
\frac{h^2}{2 v^2} \, 
G_{\rm \mu\nu}^a G^{a \, \mu\nu} \,
% \sum_i \frac{g_s^2}{192 \pi^2} C_i \left( 
\sum_i \frac{\alpha_s}{48 \pi} C_i \left( 
\frac{\kappa_i v^2}{M_i^2} - \frac{\kappa_i^2 v^4}{M_i^4} \right)
\, . \label{eq:twoHiggseff}
\end{eqnarray}
The top quark contributions to the effective couplings can
also be obtained by expanding Eq.~(\ref{eq:eff}), 
\begin{eqnarray}
% \frac{g_s^2}{48 \pi^2} \left( \frac{h}{v} - \frac{h^2}{2 v^2} \right) 
\frac{\alpha_s}{12 \pi} \left( \frac{h}{v} - \frac{h^2}{2 v^2} \right) 
G_{\rm \mu\nu}^a G^{a \, \mu\nu} \, . \label{eq:topeff}
\end{eqnarray}

There are two critical observations we can make at the
effective interaction level.  First, when $M_i > \sqrt{|\kappa_i|} v$, 
the relative \emph{sign} of the the top quark contribution to the 
effective operator $h^2 G^a_{\mu\nu} G^{a,\mu\nu}$ is is opposite 
(the same as) that of the scalar contribution when $\kappa_i > 0$ 
($\kappa_i < 0$).  Second, in the limit $\mh, v \ll M_i$, 
we see that the same coefficient that determines the single 
Higgs effective interaction also uniquely determines the 
di-Higgs effective interaction.  This will be important 
for gaining some qualitative understanding of our 
di-Higgs cross section results.

%------------------------------------------------------------------------------
\section{Contributions to di-Higgs Production}
\label{sec:strisbox}

We now carry out the calculation of di-Higgs production at a
hadron collider.  Light quarks play no role in the leading-order 
di-Higgs production cross section, so we are left with the 
gluon-induced partonic process $gg \ra hh$. 
We calculate the amplitudes for di-Higgs production in two
independent methonds.  First we calculate the amplitudes using the
effective interactions derived in Sec.~\ref{sec:let}.  
This allows us to obtain simple, analytic formulae that provide
several qualitative features of the full calculation.  
Next, we carry out a full momentum-dependent one-loop calculation
of $gg \ra hh$ including top quarks and an arbitrary set of 
colored scalars.

The amplitude for $gg \ra hh$ can be decomposed into two 
non-interfering Lorentz structures 
following Ref.~\cite{Glover:1987nx},
\begin{eqnarray}
\mathcal{M}(g_a g_b \ra hh)_{\mu\nu} &=& 
  \AP \, P_{\mu\nu} \delta_{ab} + \AQ \, Q_{\mu\nu} \delta_{ab} \; , 
\end{eqnarray}
where
\begin{eqnarray}
P_{\mu\nu} &=&   \eta_{\mu\nu} 
    - \frac{ p_{1\,\nu}\,p_{2\,\mu}}{(p_1 \cdot p_2)}\\
Q_{\mu\nu} &=&   \eta_{\mu\nu} 
    + \frac{m^2_H\, p_{1\,\nu}\,p_{2\,\mu}}{p^2_T\, p_1 \cdot p_2} 
    - \frac{2\,(p_1\cdot k_1)\,p_{2\,\mu}k_{1\,\nu}}{p^2_T\, p_1\cdot p_2} 
      \nonumber \\
& &{} \qquad 
    - \frac{2\,(p_2\cdot k_1)\,p_{1\,\nu}k_{1\,\mu}}{p^2_T\, p_1\cdot p_2} 
    + \frac{2\, k_{1\,\mu}\,k_{1\,\nu}}{p^2_T} \; .
\end{eqnarray}
In these formula, $p_1$ and $p_2$ are the incoming momenta of the gluons, 
$k_1$ is the outgoing momenta of one of the Higgs bosons, and $p_T$ 
is the transverse momenta of the Higgs: 
$p^2_T = (\hat u\,\hat t - m^4_H)/\hat s$. 
The $Q_{\mu\nu}$ is the only possible alternative Lorentz structure 
once the Ward identities and orthogonality to $P_{\mu\nu}$ 
have been imposed. 

\subsection{Di-Higgs Amplitude from Effective Couplings}

\begin{figure}[t!]
\centering
\includegraphics[width=0.45\textwidth]{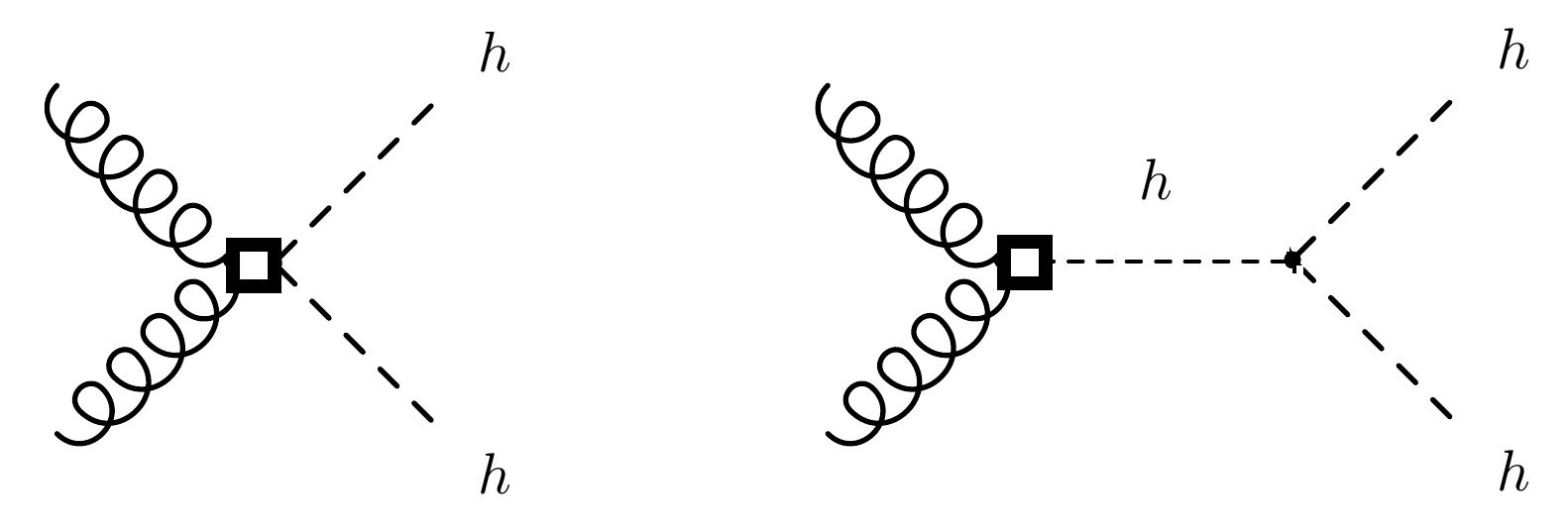}
\caption{Effective operator contributions to $gg \ra hh$.}
\label{fig:feyneff}
\end{figure}

The effective couplings 
given by Eqs.~(\ref{eq:oneHiggseff}), (\ref{eq:twoHiggseff}), 
and (\ref{eq:topeff}) lead only to contributions to the 
$P_{\mu\nu}$ structure.  
The Feynman diagrams include both the 4-point effective interaction 
$h^2\, G^a_{\mu\nu} G^{a,\mu\nu}$ as well as the diagram 
with $s$-channel Higgs exchange involving the 3-point 
effective interaction $h\, G^a_{\mu\nu} G^{a,\mu\nu}$ and the 
triple-Higgs self-interaction, shown in Fig.~\ref{fig:feyneff}.
We obtain
\begin{eqnarray}
%  1/12 x 4 = 1/3
\AP^{\rm eff}_{\rm top} &=& \frac{\alpha_s}{3 \pi v^2} 
  \left( - 1 + \frac{3 m_h^2}{\hat{s} - m_h^2} \right) \label{eq:topampeff} \\
%  1/96 x 4 = 1/24
\AP^{\rm eff}_{\rm scalar} &=& 
  \sum_i \frac{\alpha_s \kappa_i C_i}{24 \pi M_i^2} 
  \left( 1 
         + \frac{3 m_h^2}{\hat{s} - m_h^2} 
         - \frac{\kappa_i v^2}{M_i^2} \right), \label{eq:scalarampeff} 
\end{eqnarray}
where we have neglected the width of Higgs in the $s$-channel
propagators.  The top quark contributions to the amplitude
interfere destructively.  This, combined with the necessity to 
sample parton distribution functions at larger $x$ 
(to obtain larger $\hat{s}$), is the central reason that 
the di-Higgs production cross section at a hadron collider
is so small in the Standard Model.  

The amplitudes derived from the effective interactions allow us 
to make several interesting qualitative observations about the
scalar contributions:
\begin{itemize}
\item For $\kappa_i > 0$ with $M_i \gsim \sqrt{|\kappa_i|} v$,
the top contribution and the scalar contributions destructively
interfere, weakening the effects of colored scalars
on di-Higgs production.
\item For $\kappa_i < 0$ with $M_i \gsim \sqrt{|\kappa_i|} v$,
the top contribution and the scalar contributions constructively
interfere, strengthening the effects of colored scalars
on di-Higgs production.
\item When $M_i \lsim \sqrt{|\kappa_i|} v$, 
the third contribution in Eq.~(\ref{eq:scalarampeff})
begins to affect the scalar amplitude for di-Higgs production.
However, momentum-dependent corrections proportional to
$\hat{s}/M_i^2$ are at least as important, since 
$\hat{s} \ge 4 \mh^2$.  In this region, the $1/M_i$ expansion
is no longer valid, and we need the full momentum-dependent 
loop functions to make accurate quantitative calculations.  
\end{itemize}

The two main results we find from the di-Higgs amplitudes 
calculated with the effective operators are that:\,i.)
we expect a considerably larger cross section when $\kappa_i < 0$, 
and $M_i \gsim \sqrt{|\kappa_i|} v$; ii.) 
we expect a strong correlation between single Higgs production
and di-Higgs production once $M_i$ is large enough for
the effective operators to reproduce the full momentum-dependent
one-loop results.

\subsection{Di-Higgs Amplitude at One-Loop}

We now turn to the full one-loop leading-order calculation
of di-Higgs production including the top quark and the scalars. 
The Standard Model contribution to $gg \ra hh$ comes from 
top quark triangle diagrams stitched to a triple-Higgs vertex 
via a Higgs boson propagator, as well as box diagrams with 
two top Yukawa coupling insertions. The full momentum-dependent 
expressions for $\AP, \AQ$ for the top contributions can be found 
in Ref.~\cite{Glover:1987nx}.

The scalar octet loops can be similarly classified into triangles 
and boxes (or by powers of the Higgs-portal coupling $\kappa$). 
The complete set of (leading order) diagrams are shown below in 
Fig.~\ref{fig:feyn}.
\begin{figure*}[t!]
\centering
\includegraphics[width=\textwidth]{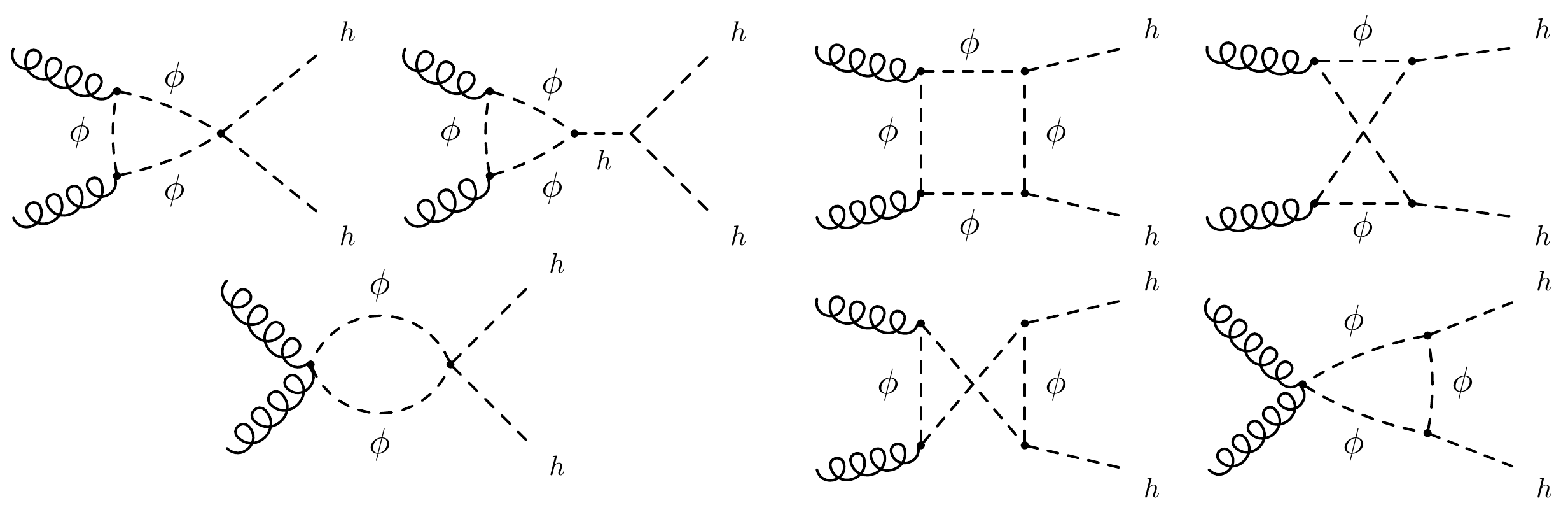}
\caption{Scalar loop contributions to $gg \ra hh$. 
The three diagrams on left are $\mathcal{O}(\alpha_s \kappa)$, while
the four diagrams on the right are $\mathcal{O}(\alpha_s \kappa^2)$, 
where $\kappa$ is the Higgs-portal coupling.}
\label{fig:feyn}
\end{figure*}
The scalar triangle diagram contributions only have 
$P_{\mu\nu}$ gauge structure:
\begin{eqnarray}
\AP_{\rm tri} &=& \sum_i
  \frac{\alpha_s \kappa_i C_i}{4\pi} 
  \left( 1+ \frac{3\, \mh^2}{\hat{s} - \mh^2} \right) \times \nonumber \\
& &{} \qquad \left( 2\, M^2_i\, C_0(p_1, p_2: M_i) + 1 \right) \, .
\label{eq:aptri}
\end{eqnarray}
The first term in the parenthesis comes from attaching the 4-point 
vertex in the $\kappa$ interaction to the scalar loop, while the
second term comes from connecting a Higgs propagator 
and 3-point vertex to the triple-Higgs self-interaction.

The box diagrams involving scalars (as well as top quarks) contribute 
to both ($P_{\mu\nu}$, $Q_{\mu\nu}$) Lorentz structures. 
We evaluate the scalar contribution to $\AP$ and $\AQ$ 
(as well as $\AP_{\rm tri}$ in Eq.~(\ref{eq:aptri}))
in terms of the Passarino-Veltman one-loop functions
given in Appendix~\ref{sec:pv}.  We obtain:
\begin{widetext}
\begin{eqnarray}
\AP_{\rm box} &=& 
  \sum_i \frac{\alpha_s\,v^2\,\kappa_i^2\, C_i}{2\pi} 
  \times \bigg[ 
      \frac{\mh^2 -\hat t}{\hat s} C_0(p_1, k_1: M_i) 
      + \frac{\mh^2 -\hat u}{\hat s} C_0(p_2, k_1: M_i)  \nonumber \\ 
& &{} \quad + M_i^2\, 
        \big( D_0(p_1, p_2, k_1: M_i) + D_0(p_2, p_1, k_1: M_i) \big)
      + \frac{\hat u\, \hat t + 2\,M^2_i\,\hat{s} - \mh^4}{2\hat s} 
        D_0(k_1, p_1, p_2 : M_i) 
  \bigg] \, , \label{eq:apbox} \\
\AQ_{\rm box} &=& 
  \sum_i \frac{\alpha_s\,v^2\,\kappa_i^2\, C_i}{2\pi} 
  \times \frac{1}{2(\mh^4 - \hat t\,\hat u)} \bigg[ 
    \hat s\,(\hat u + \hat t)\, C_0(p_1,p_2: M_i) 
     - (\hat u^2 + \hat t^2 - 2\, \mh^4)\,C_0(k_1, k_2: M_i) \nonumber \\
& &{} \quad + 2\,\hat t\,(\hat t - \mh^2)\,C_0(p_1, k_1: M_i) 
      + 2\,\hat u\,(\hat u- \mh^2)\,C_0(p_2,k_1: M_i) 
      + 2\,M_i^2\,(\mh^4 -\hat t\,\hat u)\,D_0(k_1, p_1, p_2: M_i) \nonumber \\
& &{} \quad + 
      \left( 2\,M_i^2\,(\mh^4 - \hat t\,\hat u) - \hat s\, \hat t^2 \right)
        D_0(p_1, p_2, k_1: M_i)  
      + \left( 2\,M_i^2\,(\mh^4 - \hat t\,\hat u) - \hat s\, \hat u^2 \right)
        D_0(p_2,p_1, k_1: M_i) \bigg] \, . \label{eq:aqbox}
\end{eqnarray}
\end{widetext}

Adding the scalar loop $\AP, \AQ$ to the top loop contributions 
gives us the total $\mathcal{M}(gg \ra hh)_{\mu\nu}$ amplitude. 
Squaring and adding phase space, color- and spin-averaging factors, 
we arrive at the differential partonic cross section,
\begin{eqnarray}
\frac{d\hat{\sigma}(gg \ra hh)}{d\,\cos{\theta^*}} &=& 
  \frac{\beta_h}{1024\,\pi \hat s }
  \left( |\AP_{\rm tot}|^2 + |\AQ_{\rm tot}|^2 \right) \, .
  \label{eq:gghhtot}
\end{eqnarray}
Note that the overall factor of $1/2$ for creating a pair of 
identical particles is canceled by 
$P_{\mu\nu}P^{\mu\nu} = Q_{\mu\nu}Q^{\mu\nu} = 2$.

%------------------------------------------------------------------------------
\section{Benchmark color octet model}
\label{sec:octetmodel}

We now specialize our results to our benchmark model: 
a single real, color-octet, electroweak neutral scalar 
$S_a$~\cite{Bai:2010dj,HUTP-91-A009,Dobrescu:2007yp,Dobrescu:2011aa},
\begin{eqnarray}
\mathcal{L}_S &=& \frac{1}{2} (D_{\mu} S_a)^2 - \frac{1}{2} M^2_S\, (S_a)^2 
                  - \frac{\kappa}{2} (S_a)^2\, H^{\dag}H \nonumber \\
& &{} \quad - \frac{\omega}{4} (S_a)^4 - \mu_S\, d^{abc}\,S_aS_bS_c \; .
\label{eq:lagrangian}
\end{eqnarray}
Here we have included one Higgs-portal interaction, with coupling $\kappa$, 
as well as additional renormalizable self-interactions among the $S_a$:  
a quartic interaction with coupling $\omega$, and a cubic 
interaction with (dimensionful) coupling $\mu_S$. 
The cubic interaction implies the $S_a$ can decay into two gluons 
through a triangle loop.  Since the Lagrangian possesses a $\mathcal Z_2$ 
symmetry when $\mu_s = 0$, small values of $\mu_S$ are 
technically natural.  This allows us to assume that $\mu_S$ 
is large enough to cause prompt $S_a$ decays, but is 
otherwise small enough to not play a role in our 
calculations of di-Higgs production.

The effect on single Higgs production of light scalars through 
the Higgs-portal interaction was precisely the subject of 
our Ref.~\cite{Dobrescu:2011aa}.  There we showed that 
single Higgs production could be dramatically reduced
for a wide range of \emph{negative} $\kappa$, and scalar masses 
in the range $M_S \sim 100$-$300$~GeV\@.
The reduction comes from destructive 
interference between the top quark loop and the loop of $S_a$.
The suppression could reach $75-90\%$ without excessive tuning. 
For the most extreme cancellation, relatively large couplings 
$\kappa \sim -1$ and light scalars $M_S \lesssim 200$~GeV 
were required.  Taking even lighter scalar masses, the scalar loop
contributions can overwhelm the top quark contribution, and the 
single Higgs production cross section can again be 
$\mathcal{O}(1)$ times the Standard Model value.  

As was discussed in Ref.~\cite{Dobrescu:2011aa}, color octet 
scalars are dominantly pair-produced and decay (via a loop of $S_a$)
to a pair of gluons. Thus, the signature is two pairs of jets 
with equal invariant mass, making four total jets. This signature
provides a handle to help distinguish it from the multi-jet background.
However, when the scalars are light, they create relatively soft jets
that are not efficiently triggered on due to the large background. 
In addition to the large rate and combinatorial hurdles, 
extracting the signal is further complicated by pileup.
A comparison between ATLAS~\cite{Aad:2011yh} and CMS~\cite{CMSoctets} 
multi-jet searches demonstrates the difficulties in placing 
constraints in higher luminosity runs.
The ATLAS~\cite{Aad:2011yh} search was able to disfavor real color octet 
scalar between $100$-$125$~GeV\@.  The low reach in $M_S$ was possible 
because this study was performed with $36\,\text{pb}^{-1}$ 
of data where the luminosity was sufficiently low that trigger thresholds 
could be kept small. 
The CMS~\cite{CMSoctets} search, by contrast, was performed
on $1$~fb$^{-1}$ of data, where the jet trigger threshold was 
$70$~GeV\@.  This prevented the CMS study to be sensitive to 
objects lighter than 
$300$~GeV\@.\footnote{The jet $p_T$ cut was $150$~GeV in this study.}
By adopting specialized pre-scaled triggers at lower jet $p_T$, 
improvements to the low mass octet search may be possible, 
but to the best of our knowledge, no such studies have been 
completed.

It is also important to consider the possible range of the
dimensionless Higgs-portal coupling $\kappa$, particularly for
negative values.  Large negative $\kappa$ could destabilize the 
scalar potential at large field values.  The naive tree-level 
bound can be obtained by rewriting the quartic interactions
of the scalar potential as 
$(\sqrt{\omega} (S_a)^2 - \sqrt{\lambda_h} H^\dagger H)^2 
+ 2 \sqrt{\omega \lambda_h} S_a^2 H^\dagger H$, 
and thus $|\kappa| < 2 \sqrt{\omega \lambda_h}$~\cite{Boughezal:2010ry} 
where $\lambda_h$ is the Higgs quartic coupling. 
However, in Appendix~\ref{sec:rgimp} we calculate the renormalization 
group equations for $\kappa$, $\omega$, and $\lambda_h$, showing
that negative $\kappa$ runs rapidly to smaller
(absolute) values at a larger renormalization scale.  
This suggests that the bounds on 
negative $\kappa$ are expected to be significantly relaxed 
once one uses the renormalization-group improved potential.

%------------------------------------------------------------------------------
\section{Enhanced Di-Higgs Production}
\label{sec:dihiggs}

We now calculate the di-Higgs production cross section
for two distinct scenarios:  $\mh = 125$~GeV, and $\mh > 125$~GeV\@.
In the case $\mh = 125$~GeV, we consider both negative and
positive $\kappa$, since a wide range of $(\kappa, M_S)$ parameter space
is allowed by a (generous) interpretation of the $125$~GeV
resonance observed by LHC as a Higgs boson.  
For $\mh > 125$~GeV, we consider only negative $\kappa$, 
since this is the only region of $(\kappa, M_s)$ space where
the (heavy) single Higgs production can be
suppressed below the LHC bounds \cite{Dobrescu:2011aa}.

\subsection{$\mh = 125$~GeV}
\label{sec:dihiggs125}

\begin{figure*}[!t]
\hspace*{0.55\textwidth} {\Large (b)} \\
\hfill \includegraphics[width=0.48\textwidth]{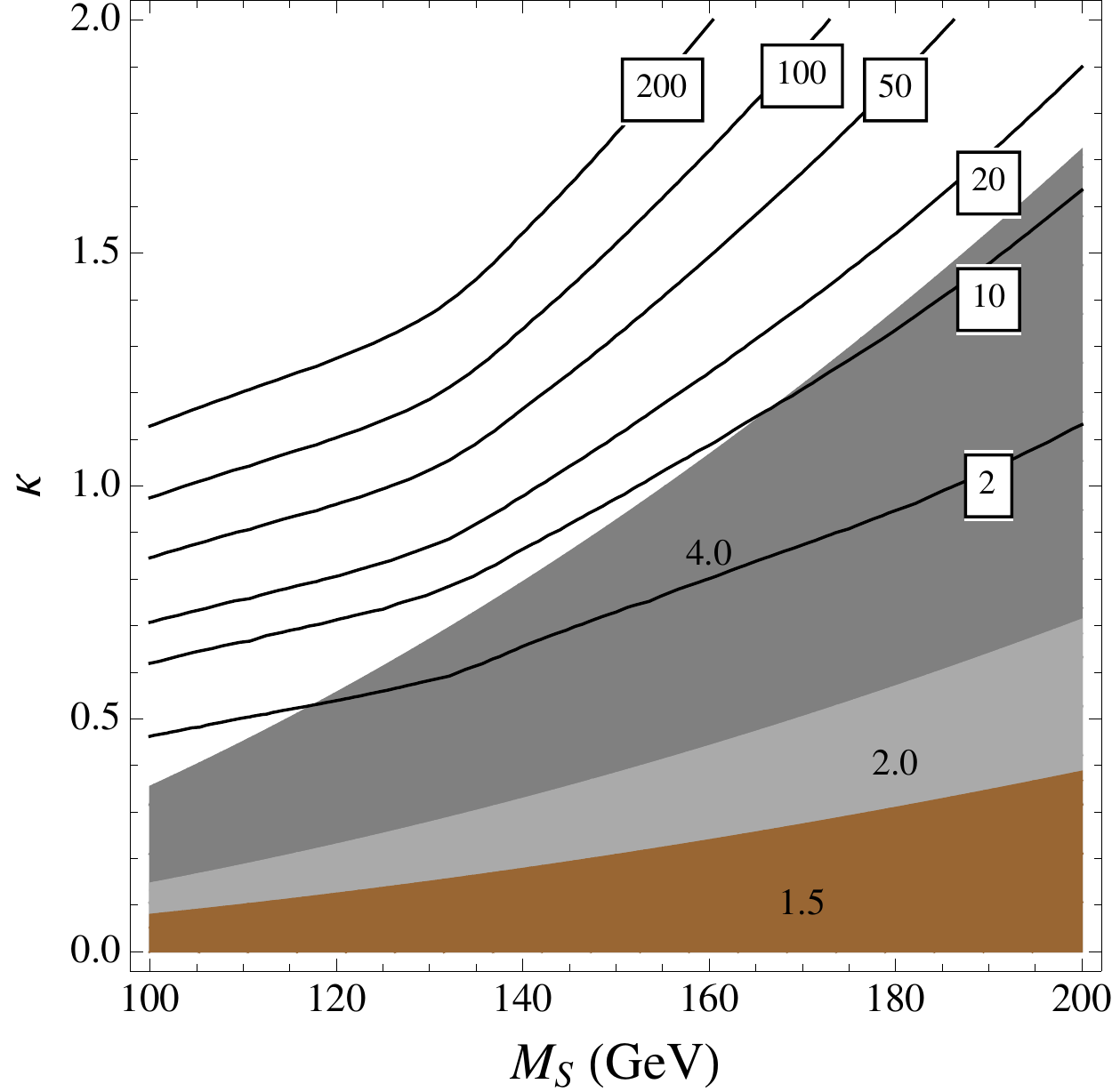} \\
\vspace*{-0.29\textwidth}
\begin{flushleft}
\hspace*{0.24\textwidth} {\Large (a)} \\
\includegraphics[width=0.49\textwidth]{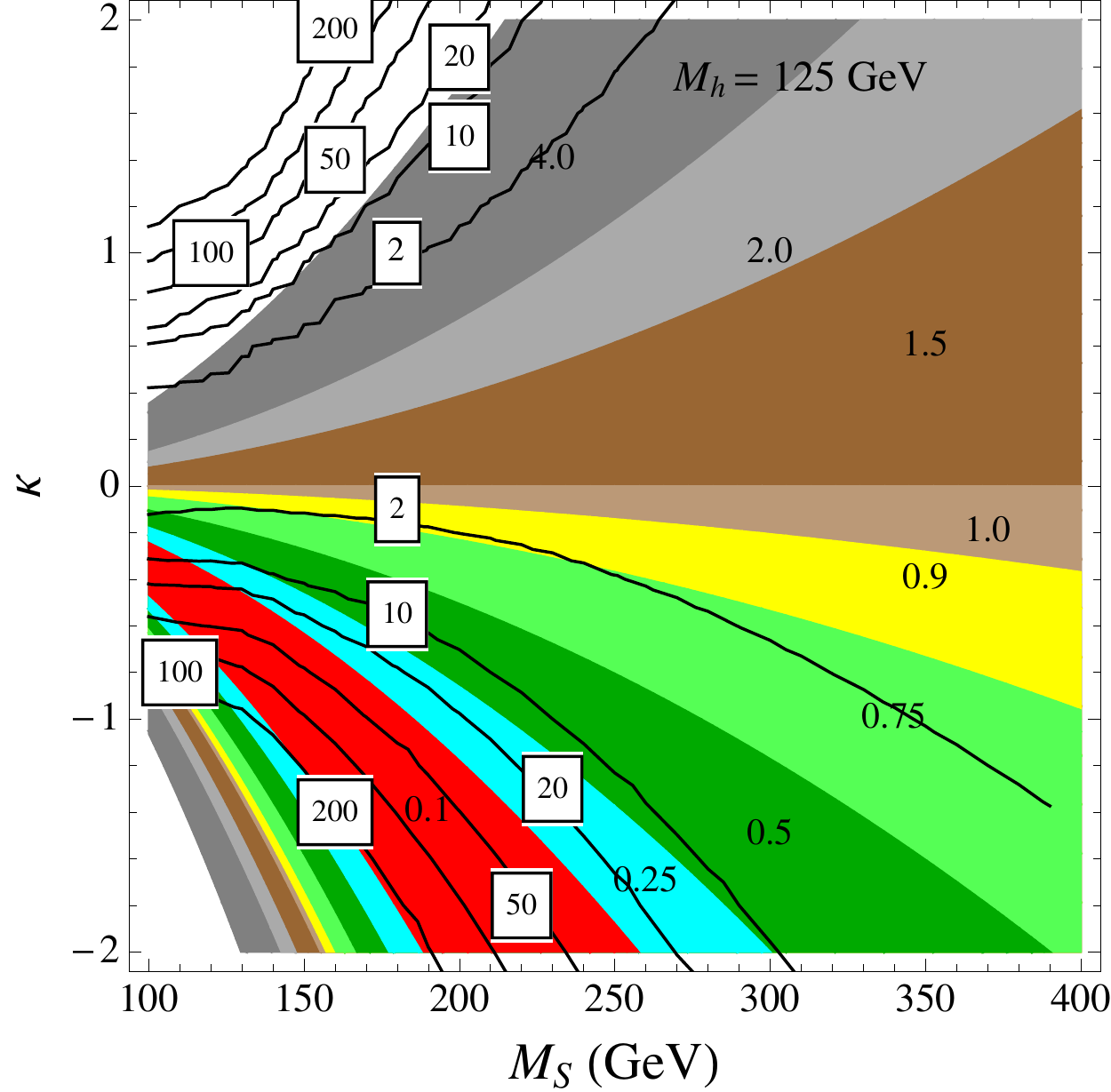} \\
\end{flushleft}
\vspace*{-0.24\textwidth}
\hspace*{0.57\textwidth} {\Large (c)} \\
\hfill \includegraphics[width=0.49\textwidth]{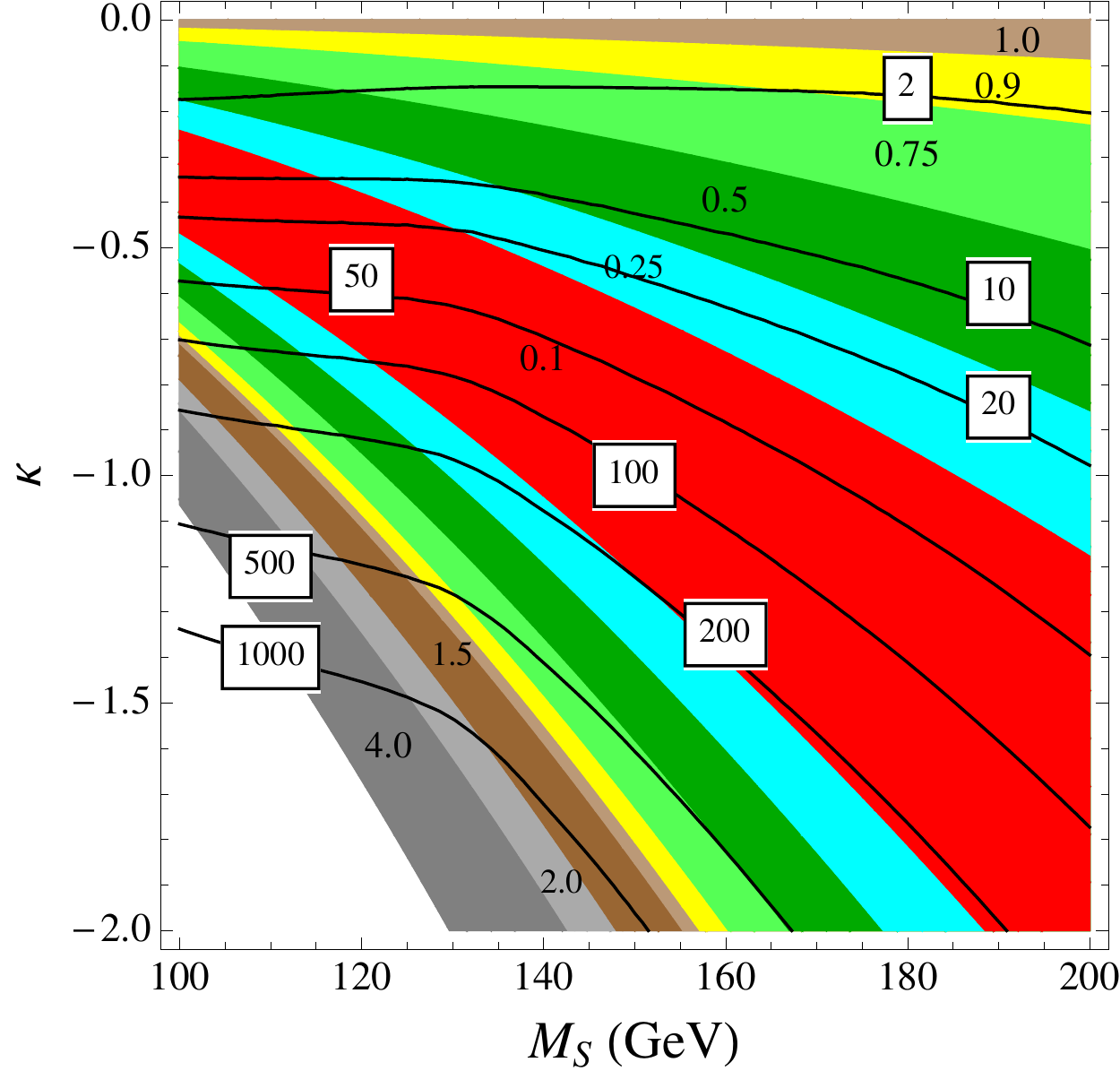} 
\caption{Di-Higgs as well as single Higgs production cross sections
normalized to the Standard Model values.  The solid lines show
$\sigma(pp \ra hh)_{\rm scalar+top}/\sigma(pp \ra hh)_{\rm top}$,
while the colored regions show when 
$\sigma(pp \ra h)_{\rm scalar+top}/\sigma(pp \ra h)_{\rm top}$
is less than the numerical value labeling each region.
Panels (b) and (c) are ``zoomed in'' versions of panel (a), 
to more clearly see the contours in the light $M_S < 200$~GeV 
region.}
\label{fig:125}
\end{figure*}

We are finally in a position to evaluate the di-Higgs production
cross section, Eq.~(\ref{eq:gghhtot}), 
comparing to the Standard Model value while scanning over the 
scalar parameters $M_S$ and $\kappa$.  In Fig.~\ref{fig:125} 
we show numerical results for the ratio
$\sigma(pp \ra hh)_{\rm top + scalar}/\sigma(pp \ra hh)_{\rm top}$
as a function of $M_S$ and $\kappa$ for LHC at $\sqrt{s} = 8$~TeV\@.
The scalar and top calculations were performed at leading order;
however by taking a ratio to the Standard Model result, 
higher order corrections in $\alpha_s$ to di-Higgs production 
should largely cancel out.  Fig.~\ref{fig:125} contains three plots: 
one with positive and negative $\kappa$ with $M_S < 400$~GeV; 
and two ``zoomed in'' plots that more easily demonstrate the
large effects at smaller $M_S$ for both positive and 
negative $\kappa$.
These contours are overlaid onto colored regions that show 
the single Higgs production ratio
$\sigma(pp \ra h)_{\rm top + scalar}/\sigma(pp \ra h)_{\rm top}$,
derived from Ref.~\cite{Dobrescu:2011aa}.
We see that once $M_S \gsim \mh$,
the di-Higgs production contours follow the single Higgs production
region shapes fairly well.  This was partly anticipated from the
coefficients of the effective operators, 
Eqs.~(\ref{eq:oneHiggseff}),(\ref{eq:twoHiggseff}), 
that were found to be the same for $h/v G^a_{\mu\nu} G^{a,\mu\nu}$ 
and $h^2/(2 v^2) G^a_{\mu\nu} G^{a,\mu\nu}$.
Interestingly, the \emph{magnitude} of the
di-Higgs production cross section is not estimated particularly
well from the effective operators.  
In Appendix~\ref{sec:operatorcompare} we compare the results 
for the exact momentum-dependent one-loop calculation and
the effective effective operators.  This we do by showing
the negative $\kappa$ ``zoomed in'' plot identical to
Fig.~\ref{fig:125}(c) overlaid with the effective operator 
results.  It is perhaps not surprising that these results
do not agree, since the effective operators neglect
momentum-dependent corrections of order $\hat{s}/M_i^2$
that is order one or larger in the region we show.
Perhaps more interesting, however, is that the effective
operators do reproduce the \emph{shape} of the contours
of the di-Higgs cross section ratio in $(\kappa, M_S)$
when the ratio 
$\sigma(pp \ra hh)_{\rm scalar+top}/\sigma(pp \ra hh)_{\rm top}$
itself is large (say, $\gsim 10$)
and $M_S \gsim \mh$.  Of course in the region 
$M_S < \mh$, as well as when the ratio itself is 
relatively small (say, $\lsim 10$), the effective operators
do not reproduce either the magnitude or shape of the
cross section ratio results.  

If we generously allow the recent LHC results to suggest
$0.5 < \sigma(pp \ra h)_{\rm scalar + top}/\sigma(pp \ra h)_{\rm top} < 2$,
then there are three regions of interest:  
\begin{itemize}
\item Negative $\kappa$, below and to the left of the red region.
This region has the largest di-Higgs cross section enhancement,
since it is the region with largest negative $\kappa$ for a 
given $M_S$.  In this region the scalar contributions to the
single Higgs production are roughly twice the size, but opposite
in sign, to the top contribution, resulting in a single Higgs
production rate that can be smaller, the same as, or larger than
the SM single Higgs production rate.  The di-Higgs cross section,
by contrast, is \emph{enormous} -- between several roughly
$100$-$1000$ times the Standard Model cross section.  
\item Negative $\kappa$, above and to the right of the red region.
This region has a single Higgs production cross section that
is smaller than the Standard Model.  Nevertheless, the di-Higgs
production cross section can be up to about $10$ times the
Standard Model rate.  In this region there is a strong
correlation between (slight) suppression of the single Higgs
production rate and the (much larger) enhancement of the
di-Higgs production rate.
\item Positive $\kappa$.  In this region, the single Higgs production
rate is larger than the Standard Model.  Restricting to 
$\sigma(pp \ra h)_{\rm scalar + top}/\sigma(pp \ra h)_{\rm top} < 2$,
we see the di-Higgs enhancement is negligible.  It is in this
region that smaller positive $\kappa$ leads to constructive 
interference for single Higgs production and destructive
interference for di-Higgs production.  
\end{itemize}

\subsection{$\mh > 125$~GeV}
\label{sec:dihiggsothermh}

\begin{figure*}[t]
\includegraphics[width=0.48\textwidth]{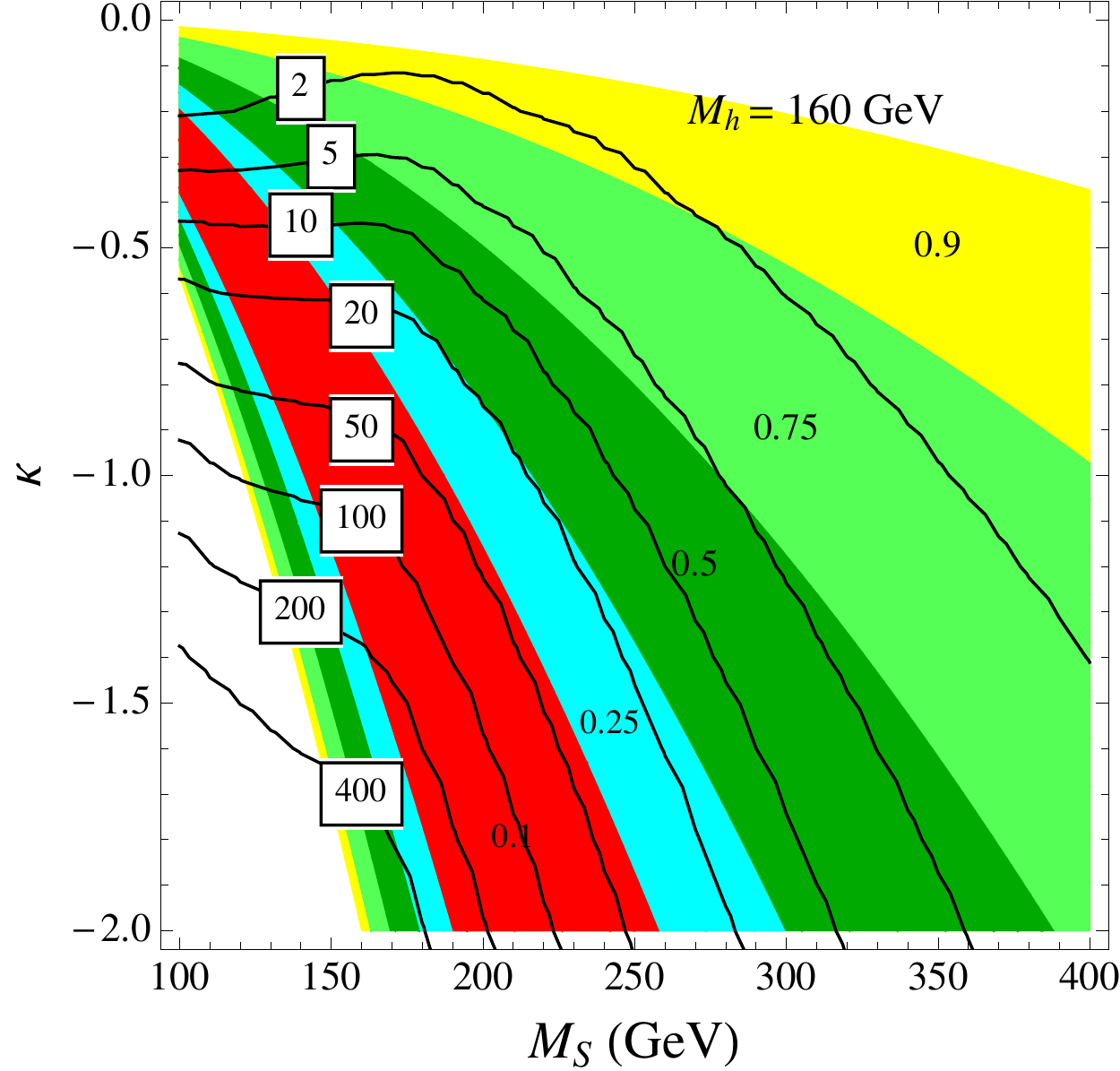}
\hfill
\includegraphics[width=0.48\textwidth]{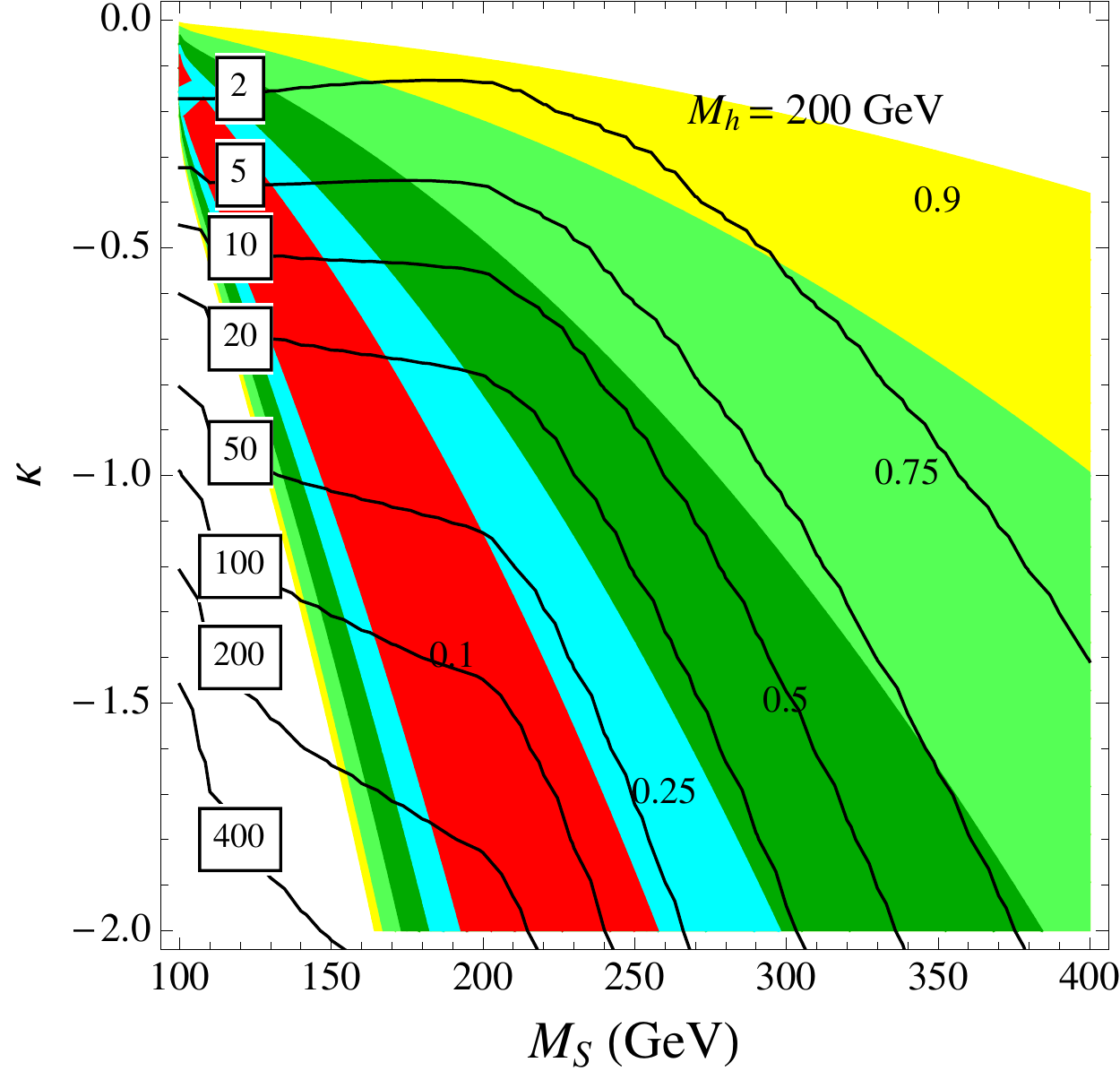} \\ \vspace{0.2in}
\includegraphics[width=0.48\textwidth]{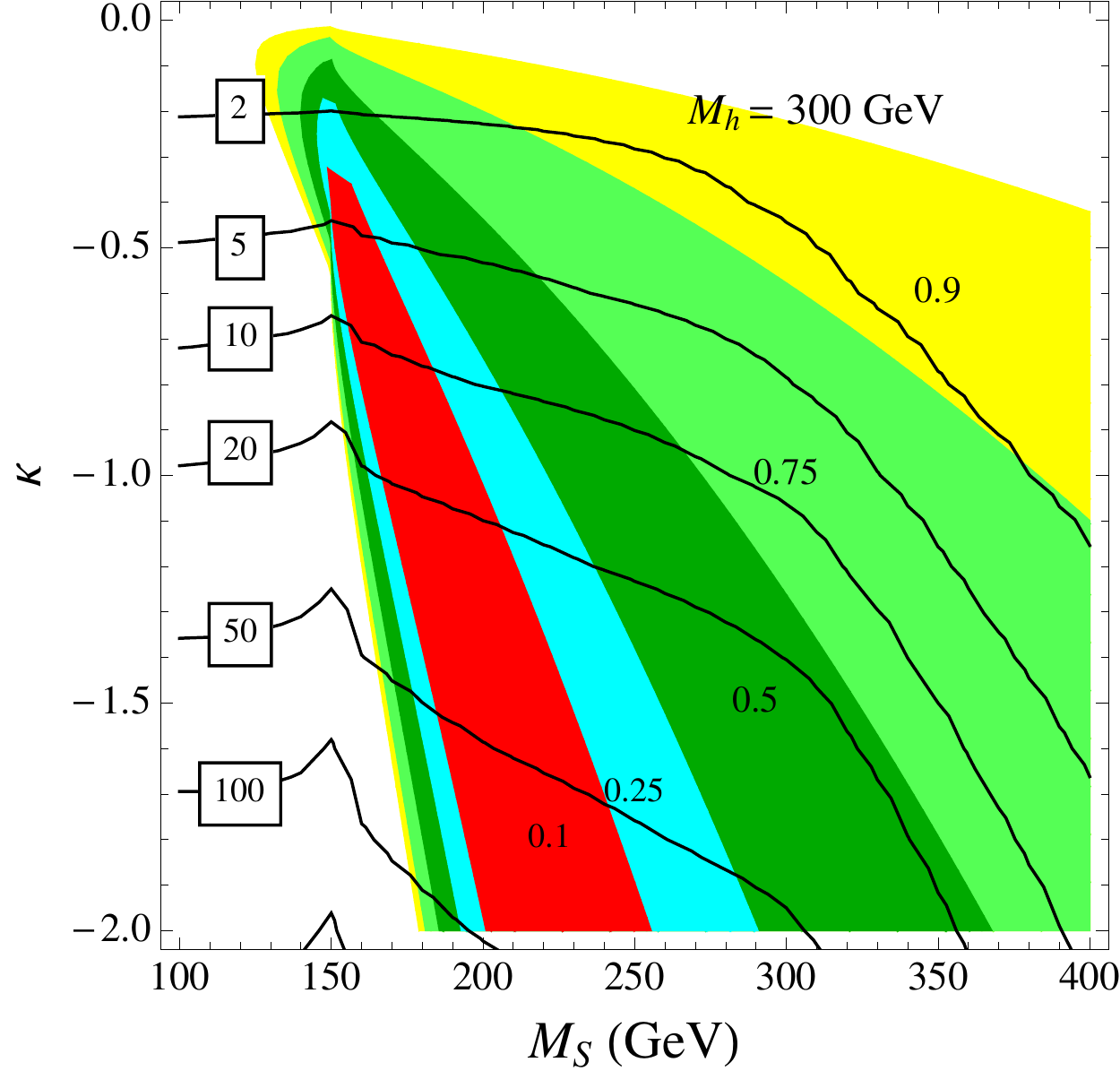}
\hfill
\includegraphics[width=0.48\textwidth]{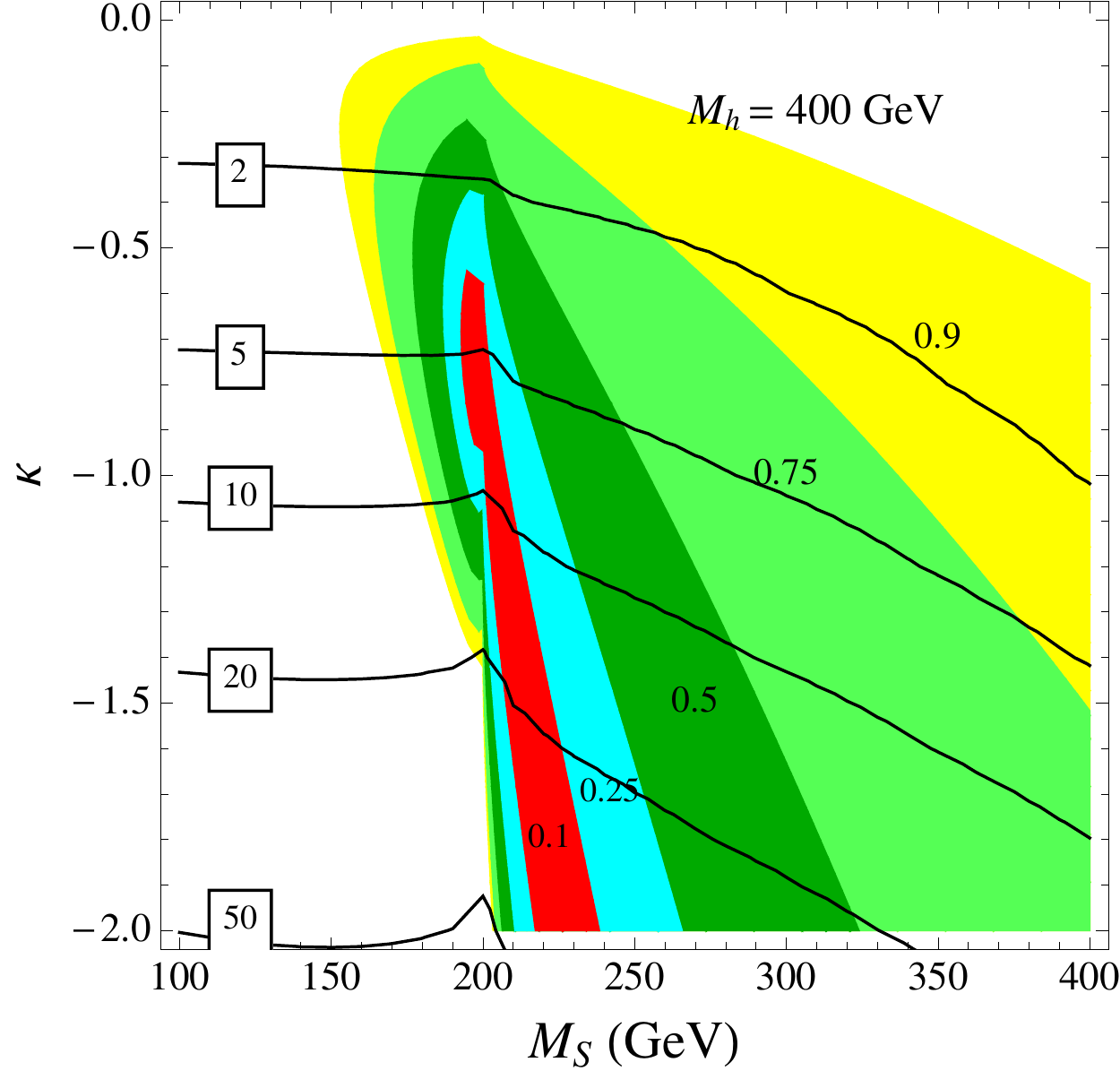}
\caption{Same as Fig.~\ref{fig:125}, but for heavier Higgs masses
(assuming the $125$~GeV particle observed by LHC is an imposter).
The viable parameter space is roughly in the red and blue regions
where 
$\sigma(pp \ra h)_{\rm scalar + top}/\sigma(pp \ra h)_{\rm top} \lsim 0.25$.
Whenever $\mh > 2\, M_S, 2\, m_t$ there is little space for a suppressed 
$\sigma(pp \ra h)$. This occurs because the loop contributions 
to the amplitude become complex, and cancellation must be arranged simultaneously between the real and imaginary parts of the top quark and scalar loops.}
\label{fig:othermh}
\end{figure*}

Consider now the case where the $125$~GeV particle observed by LHC is, 
in fact, an imposter.  As we showed in 
Ref.~\cite{Dobrescu:2011aa}, there is viable parameter region
where the Higgs can be much heavier, and yet, be safe against
the LHC search bounds on single Higgs production.
This occurs when the colored scalar contributions to single Higgs
production interfere destructively with the top contribution,
leading to a large region of highly suppressed single Higgs
production rate at LHC\@.  We show this in Fig.~\ref{fig:othermh},  
where the viable region lies entirely in the red and some of the blue region, 
where the single Higgs cross section at LHC, 
$\sigma(pp \ra h)_{\rm scalar + top}/\sigma(pp \ra h)_{\rm top} < 0.1$-$0.25$.
Here we see that there is a perfect correlation between 
\emph{suppression} of single Higgs production and 
\emph{enhancement} of di-Higgs production.  
This occurs for all of the Higgs masses shown, with the 
largest di-Higgs enhancement occurring for the smaller Higgs masses,
$\mh = 160,200$~GeV, large negative $\kappa$, and $M_S \lsim 250$~GeV\@.
Here the enhancement is between about $5$ to $200$ times
the Standard Model rate (for the given Higgs mass).  
This is the smoking gun for a heavy hidden Higgs boson in the
case where the $125$~GeV particle is an imposter.  

%------------------------------------------------------------------------------
\subsection{Kinematic Distributions of Enhanced Di-Higgs Production}
\label{sec:dihiggskin}

In addition the large increase in the di-Higgs production rate,
there are also modifications to the kinematical distributions
of di-Higgs production.  Here we turn to Monte Carlo to simulate 
the signal.  We implemented the $gg \ra hh$ processes 
into {\tt MadGraph4}~\cite{Alwall:2007st} by modifying the necessary {\tt HELAS}~\cite{Hagiwara:1990dw, Murayama:1992gi} routines, including both Lorentz structures 
($P_{\mu\nu}$ and $Q_{\mu\nu}$) and retained the full momentum dependence. 
To evaluate the Passarino-Veltman one-loop functions, 
the $gg \ra h, gg \ra hh$ amplitudes were interfaced with the 
LoopTools~\cite{Hahn:1998yk} package. We utilize CTEQ6L1 parton distribution 
functions with the scale choice $\mu_F = \mu_R  = 2\times m_h$ for all simulations.

We first look to basic kinematic distributions of the Higgs bosons.
The $p_T$ and rapidity ($Y_h$) spectra are shown below in Fig.~\ref{fig:kinematics1} 
for several choices of $M_S$; the distributions are area-normalized to 
focus on the shape difference. Note that the $p_{T,h}$ spectrum peaks 
roughly at the mass of the particle dominating the loops, 
near $m_t$ in the SM and $M_S$ in all non-SM cases. 
For the SM case this feature can be traced to an enhancement 
in the diagrams when $\hat{s} \sim m^2_t$~\cite{Glover:1987nx}. 
A similar enhancement occurs in the scalar loops at $\hat{s} \sim M_S^2$. 
However, unlike in the SM where the triangle and box diagrams 
always destructively interfere, the interference among the 
various scalar diagrams depends on the sign of $\kappa$,
as we have seen.  For positive $\kappa$, the scalar boxes and triangles 
also interfere destructively.  The severity of the interference depends 
on $|\kappa|$, since the box contributions grow as $\mathcal{O}(\kappa^2)$, 
whereas the triangles, $\mathcal{O}(\kappa)$.  

\begin{figure*}
\hspace*{0.06\textwidth} {\Large (a)} 
\hspace*{0.48\textwidth} {\Large (b)} \\
\includegraphics[width=0.48\textwidth]{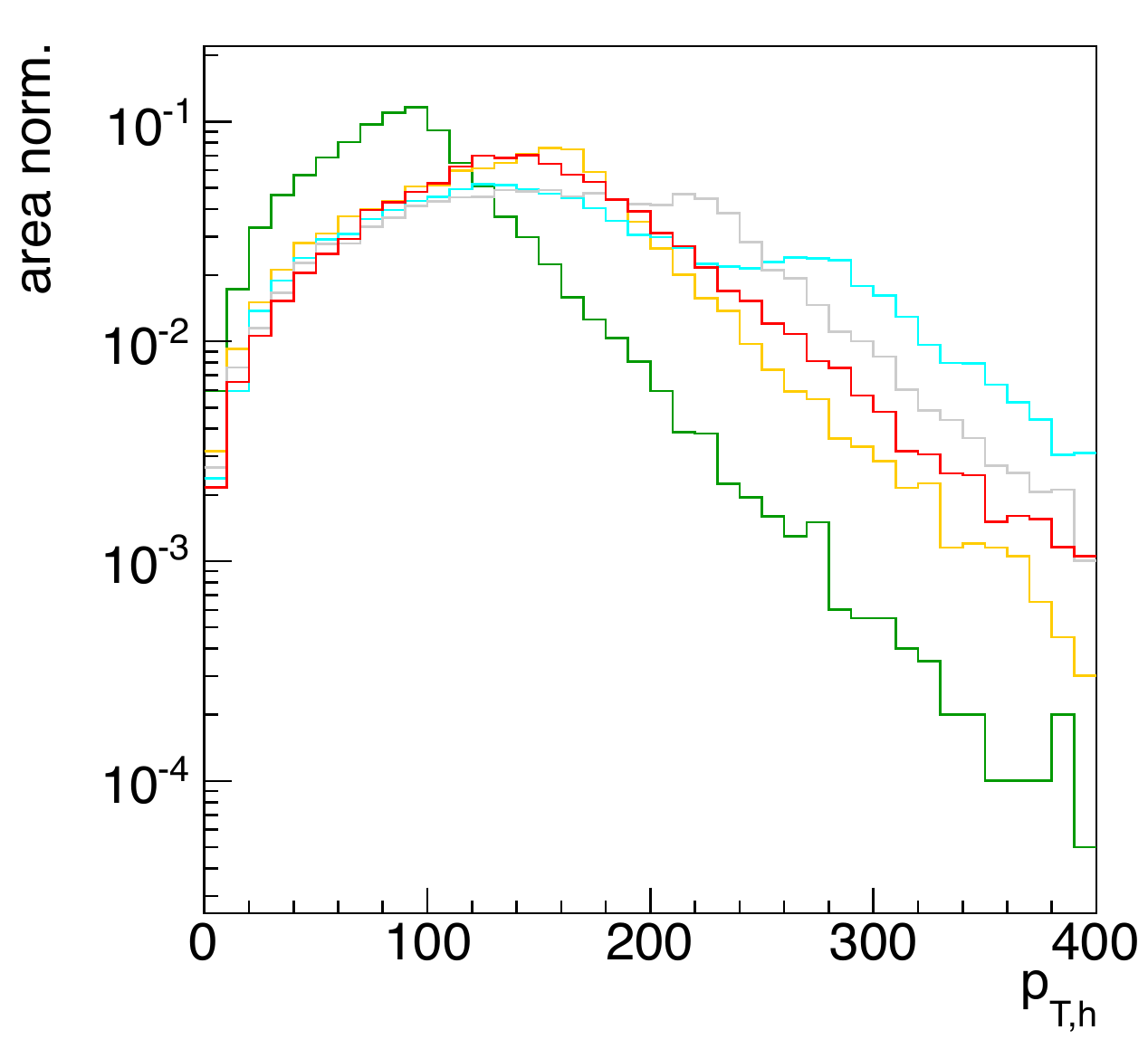}
\hfill 
\includegraphics[width=0.48\textwidth]{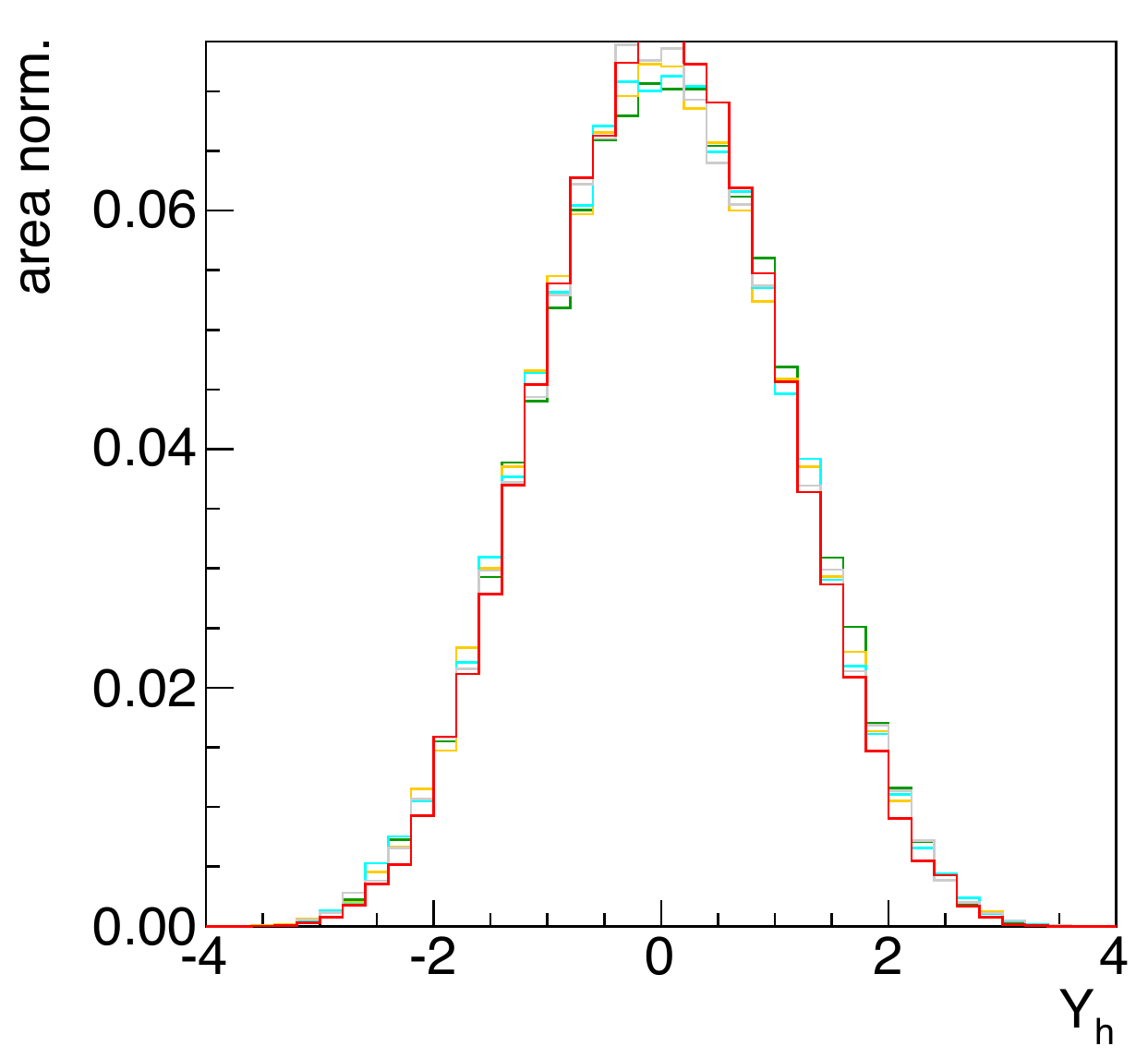}
\caption{Kinematic distributions for the $p_{T,h}$ of the
Higgs boson in di-Higgs production at LHC\@.  
Panel (a) shows $(1/\sigma) d\sigma(gg \ra hh)/dp_{T,h}$ for the 
SM alone (red) and the SM combined with loops of scalars, 
$M_S = 160\,\gev$ (green), 
$M_S = 200\,\gev$ (orange), 
$M_S = 250\,\gev$ (gray) and 
$M_S = 300\,\gev$ (cyan). 
The Higgs-portal has been taken to be $\kappa = -1$ for all cases. 
The right panel shows the Higgs rapidity distribution
$(1/\sigma) d\sigma(gg \ra hh)/d Y_{h}$ 
for the same scalar values. }
\label{fig:kinematics1}
\end{figure*}

\begin{figure}[t]
\centering
\includegraphics[width=0.48\textwidth]{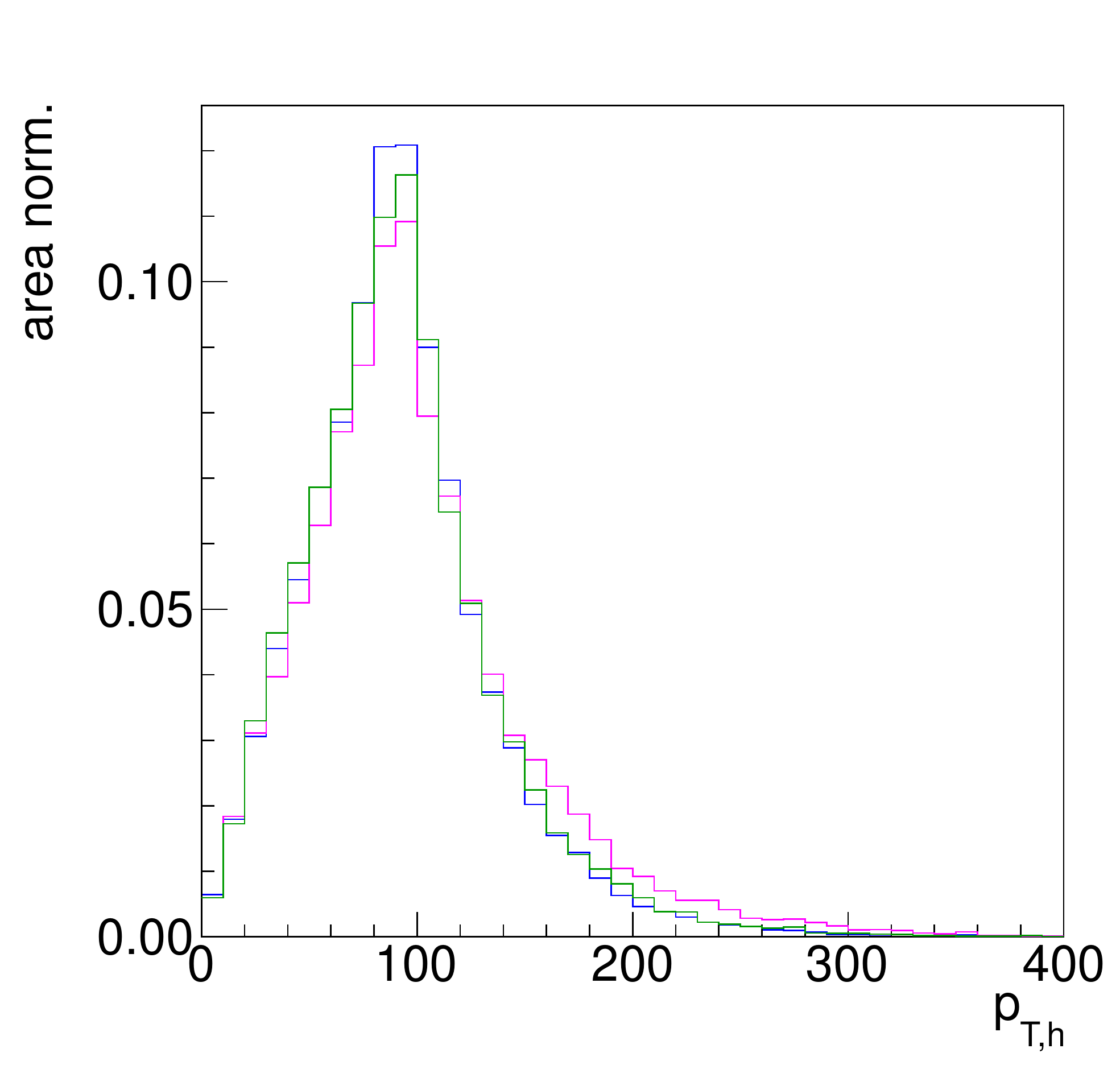}
\caption{Comparison between the $p_{T,h}$ spectra in $gg \ra hh$ 
for three different values of $\kappa$.  The scalar mass is 
held fixed at $160\,\gev$ but we vary the Higgs-portal coupling; 
$\kappa = -1.0$ (green), 
$\kappa = -1.5$ (blue), 
$\kappa = -0.5$ (magenta). 
The distributions are area normalized to focus on the shape differences.}
\label{fig:kinematics2}
\end{figure}

In Fig.~\ref{fig:kinematics1} the Higgs-portal coupling was fixed 
to $\kappa = -1$, while the scalar mass was varied.  
However, as there are both $O(\kappa)$ and $O(\kappa^2)$ 
contributions to $gg \ra hh$ (at amplitude level), 
each weighted by different kinematic functions, the $p_{T,h}$ spectrum 
does carry some $\kappa$ dependence. As we can see from Fig.~\ref{fig:kinematics2}, which shows the overlayed $p_{T,h}$ spectra for three different $\kappa$ values and fixed $M_S = 150\,\gev$, the $\kappa$ dependence
is fairly small. 

\begin{figure}
\centering
\includegraphics[width=0.48\textwidth]{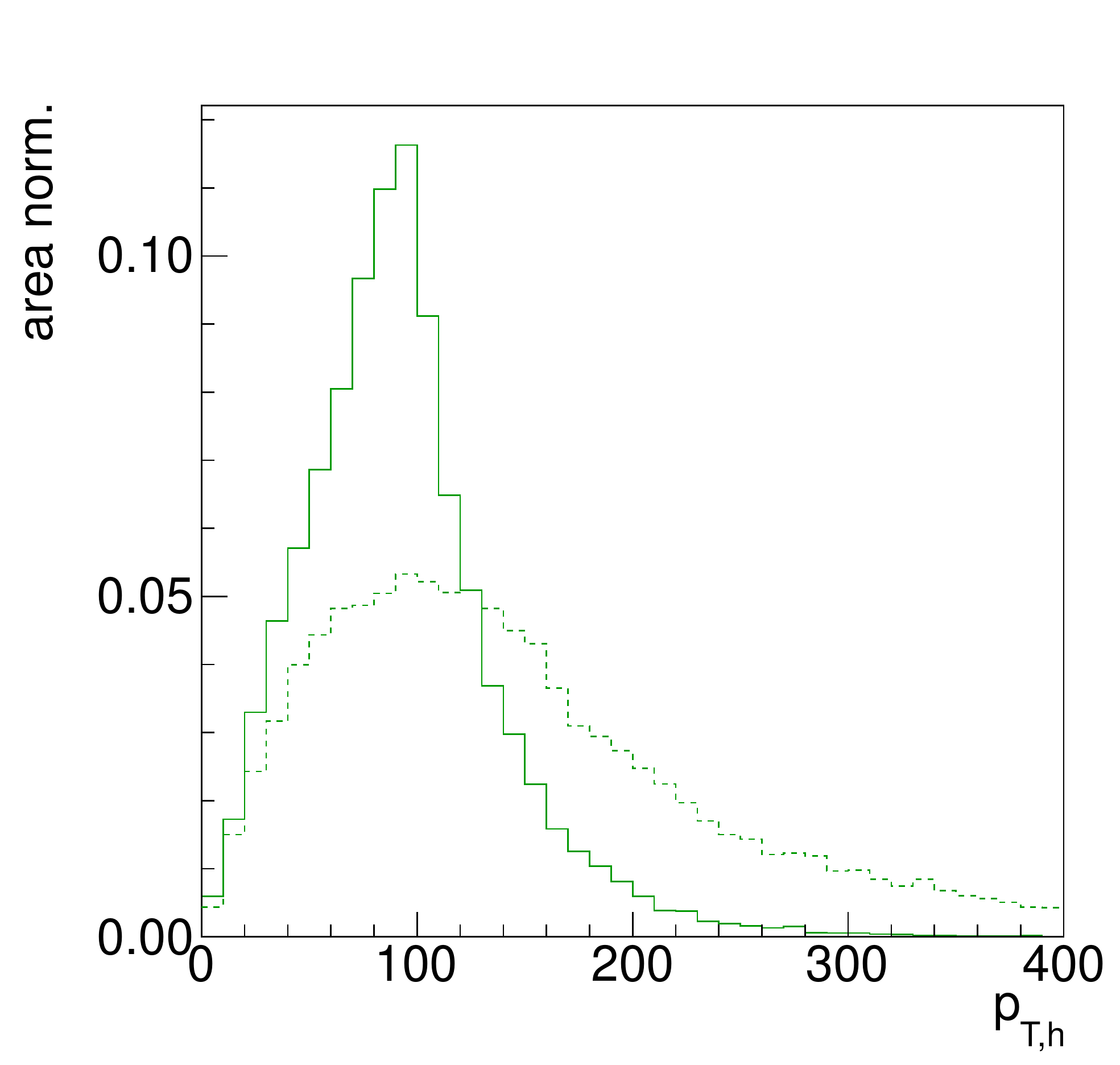}
\caption{The difference in the $p_{T,h}$ spectrum of the Higgs bosons
from di-Higgs production between using the full one-loop momentum dependence 
(solid line) versus the effective operators (dashed line).  
In this figure we have taken $M_S = 150\,\gev$, $\kappa = -1$. 
The distributions are area normalized to focus on the 
shape differences.}
\label{fig:kinematics3}
\end{figure}

Next, we come to the differences between using effective operators 
and retaining the full one-loop momentum-dependence. 
In Fig.~\ref{fig:kinematics3}, we take $M_S = 150\,\gev$, $\kappa = -1$ 
and compare the $p_{T,h}$ distributions between using the
effective operators [Eqs.~(\ref{eq:topampeff})-(\ref{eq:scalarampeff})] 
against the one-loop results [Eqs.~(\ref{eq:aptri})-(\ref{eq:aqbox})].
Clearly, the distributions are qualitatively distinct
throughout the $p_{T,h}$ range.
For the effective operator calculation, the high-$p_T$ tail is 
governed by kinematics alone, where as there is an extra suppression 
from the form factors once we incorporate momentum dependence. 
An accurate $p_T$ spectra is vital for phenomenology; the 
Higgs $p_T$ will be transferred to its decay remnants, 
and the details of the remnant kinematics are a necessary ingredient 
for successfully separating signal from SM background, 
regardless of the identity of the final state particles.

%------------------------------------------------------------------------------
\section{Discovering Di-Higgs}
\label{sec:discover}

The signals and discovery strategies for di-Higgs production crucially
depend on the mass of the Higgs boson and how much the rate is enhanced. 
We will not attempt to cover all scenarios, and instead focus on
two distinct cases:  $\mh = 125$~GeV and $\mh = 200$~GeV\@.
The $\mh = 200$~GeV scenario is fairly representative of a generic
heavier Higgs mass ($\mh \gsim 200$~GeV), since the branching fractions 
into massive gauge bosons dominate.  
We will sketch the best signals for each case and consider 
strategies to improve the discovery potential. 
A detailed study of the full background rates and the 
optimal cut strategy for a given Higgs mass is beyond the scope 
of this work.

In the case $\mh = 125$~GeV (where the particle observed by LHC is
taken to be the Higgs boson), the rate for single Higgs production is
constrained to be roughly the Standard Model value.
As we saw from Sec.~\ref{sec:dihiggs125}, given a range in the 
single Higgs production rate, there is a corresponding range
of $\kappa, M_S$ parameter space.  The wider the range, the larger
the variation in the di-Higgs production rate.  

Here we are interested in the absolute rate of di-Higgs production.
While we have not calculated beyond one-loop, Ref.~\cite{Dawson:1998py}
found the $K$ factor for di-Higgs production in the Standard Model
to be $\mathcal{O}(2)$.  This means 
%  20 x 10^3/(4 x 2) = 2500
$\sigma(pp \ra h)/\sigma(pp \ra hh) \simeq 2500$ at $\sqrt{s} = 8$~TeV\@.
Because there are two Higgs bosons in the final state, the rate for 
di-Higgs to feed into any characteristic single-Higgs final state 
($\gamma\gamma+X, ZZ^*+X$, etc.) is obviously doubled.
We expect that these di-Higgs events would be captured by the inclusive 
single Higgs searches at the LHC\@.  In the absence of an observation, 
this should allow ATLAS and CMS
to obtain an estimate of the upper bound on the di-Higgs enhancement. 
Dedicated analyses of existing $\sqrt{s}=7$ and $8$~TeV data 
that focus on the exact di-Higgs final state and kinematics, such as 
$\gamma\gamma + \bar b b$, would undoubtedly increase the sensitivity
to the signal.

The discovery prospects for some di-Higgs final states, 
including with enhanced production rate, have been outlined in 
Ref.~\cite{Baur:2002rb, Baur:2002qd, Baur:2003gp,  Binoth:2006ym,Pierce:2006dh,Contino:2012xk} . 
However, in the time since those studies there have been 
significant advances in the usage of jet substructure as a 
discovery tool, for both hadronic ($\bar b b$)~\cite{Butterworth:2008iy,Abdesselam:2010pt,Altheimer:2012mn} and 
$\tau^+\tau^-$~\cite{  Katz:2010iq, Englert:2011iz} resonances.  
Given the relatively high $p_T$ that the Higgs bosons can carry,
i.e., Fig.~\ref{fig:kinematics1}, we expect di-Higgs events to be
well suited to substructure techniques.

The rates, in $\fb$, for several potential di-Higgs final states 
are shown in Table~\ref{table:rates}. 
All rates are functions of the enhancement $X$ and we have taken 
the $K$-factor to be $2.0$ throughout. 
We find that varying the parton distribution set changes the cross-section by 
$\mathcal{O}(20\%)$.

\begin{table}[t!]
\centering
\begin{tabular}{c|c|ll} 
& final state                & \multicolumn{2}{c}{rates in fb} \\
&                            & \multicolumn{2}{c}{$8$~TeV~~~$14$~TeV} \\ 
                               \hline\hline
& $\gamma\gamma+\bar b b$    & $0.019\, X$ & $0.084\,X$ \\ 
$\mh = 125$~GeV 
& $\tau^+\tau^- + \bar b b$  & $0.53\, X$  & $2.4\, X$  \\
& $\tau^+\tau^-\tau^+\tau^-$ & $0.029\, X$ & $0.13\, X$  \\ \hline
& $W^+ W^- W^+ W^-$          & $1.21\, X$   & $5.73\, X$  \\ 
$\mh = 200$~GeV 
& $Z Z Z Z$                  & $0.14\, X$  & $0.68\, X$ \\ 
& $W^+ W^- Z Z$              &$0.42\, X$  & $1.97\, X$  \\ 
\end{tabular}
\caption{The di-Higgs signal production rates into the different modes
when $\mh = 125$~GeV and $\mh = 200$~GeV\@. 
All rates, which include a K factor of 2.0, must be multiplied by the enhancement factor $X$
which can be read off from the di-Higgs contours in 
Fig.~\ref{fig:125} ($\mh = 125$~GeV) and Fig.~\ref{fig:othermh}
($\mh = 200$~GeV).}
\label{table:rates}
\end{table}

Now, in the alternative case $\mh = 200$~GeV, the single Higgs
production cross section is constrained to be less than the
present LHC bounds.  This basically requires us to choose
$(\kappa, M_S)$ within the red or blue region of 
Fig.~\ref{fig:othermh}.  By taking $\mh = 200$~GeV, both
diboson decay modes are open, i.e., $\mh > 2\,m_W, 2\,m_Z$, 
and the di-Higgs production phenomenology becomes fairly insensitive 
to the exact value of the Higgs mass.  In Table~\ref{table:rates}
we show the rates for the various combinations of dibosons.
The $hh \ra W^+ W^- W^+ W^-$ is the largest, and provides
several opportunities involving same-sign dileptons or trileptons
plus missing energy.  Here again, jet substructure techniques 
could provide valuable additional sensitivity given the large
$p_T$ of the Higgs bosons in di-Higgs production.

%------------------------------------------------------------------------------
\section{Discussion}

We have demonstrated that di-Higgs production at the LHC
could be many orders of magnitude larger than the 
Standard Model in a the presence of light colored scalars.
The large enhancements are possible if $\mh = 125$~GeV, 
consistent with the recent data from LHC, 
as well as if $\mh > 125$~GeV, should the $125$~GeV particle
turn out to be a Higgs imposter.  The latter region is
natural in a model with light colored scalars, since this model
was already demonstrated to effectively hide the Higgs
below the LHC limits \cite{Dobrescu:2011aa}. 
The largest effects occur when the Higgs-portal coupling
$\kappa$ is negative and colored scalars that are lighter
than the electroweak breaking scale.  
Our detailed numerical
results were performed in a model with a single, real, 
color octet scalar.  Direct production of the colored scalar
and decay to pairs of gluons is presently constrained by LHC
data only for $M_S \lsim 125$~GeV\@.  Hence, a wide range of
$(\kappa, M_S)$ space remains viable that can lead to the huge
di-Higgs enhancement rates that we find.

Our calculation of the production rates were performed 
for a general set of colored scalars at leading order,
and thus up to the overall multiplicity, are independent
of the electroweak quantum numbers.  
The detailed branching fractions of the Higgs will, of course,
depend on the electroweak quantum numbers.  Generally this 
leads to $\mathcal{O}(1)$ effects on the branching fractions,
whereas the di-Higgs production rate enhancements we found
can be orders of magnitude larger.  It would be interesting to
perform a more detailed investigation of the correlation between
single Higgs production and decay, with other representations of
colored scalars, versus the di-Higgs production 
cross section~\cite{Mantry:2007ar,Arnold:2009ay,Bai:2011aa,Batell:2011pz}.
In addition, electroweak charged scalars will alter a wider set of 
processes than electroweak singlets: $pp \ra VV$, 
where $V = \gamma, Z^0, W^{\pm}$ will all change due to loops of 
electroweak charged scalars, in addition to $pp \ra hh$. 
Restricting to electroweak neutral colored scalars, we were somewhat 
forced to use octets.  This is because smaller representations, 
for instance the color triplet or sextet, are \emph{stable} when  
the scalars are electroweak singlets. 

The signals of di-Higgs production provide several exciting
opportunities for the LHC\@.  We have already seen that di-Higgs 
production could have \emph{already} contributed a fraction
of the single Higgs production rate, with the second Higgs
missed (or unidentified) in the events.  If $\mh = 125$~GeV,
among the more interesting di-Higgs signals includes
$hh \ra \gamma\gamma\bar{b}b$, $hh \ra \tau^+\tau^-\bar{b}b$,
and $hh \ra \tau^+\tau^-\tau^+\tau^-$.  Each of these signals
has two pairs of particles that reconstruct to two Higgs bosons.
If $\mh > 125$~GeV, there are several channels resulting from
$hh \ra W^+W^-W^+W^-$, including same-sign leptons plus missing
energy as well as trileptons plus missing energy that could be
seen at LHC\@.  The latter two searches might first obtain evidence
from recasted supersymmetric searches, while a more dedicated
search strategy would undoubtedly improve the sensitivity.
Jet substructure may also provide a valuable additional strategy 
for both low and high Higgs masses.

%------------------------------------------------------------------------------
\appendix

%------------------------------------------------------------------------------
\section{Passarino-Veltman One-Loop Functions}
\label{sec:pv}

The Passarino-Veltman functions are defined by:
\begin{widetext}
\begin{eqnarray}
C_0(p_i, p_j : m) &=& \int \frac{d^4 q}{i\pi}
\frac{1}{(q^2 - m^2)((q+p_i)^2 - m^2)((q+p_i +p_j)^2-m^2)} \\
D_0(p_i, p_j, p_k : m) &=& \int \frac{d^4 q}{i\pi}
\frac{1}{(q^2 - m^2)((q+p_i)^2 - m^2)((q+p_i+p_j)^2-m^2)((q+p_i+p_j+p_k)^2-m^2)}
\end{eqnarray}
\end{widetext}

%------------------------------------------------------------------------------
\section{Effective Couplings versus One-loop Results}
\label{sec:operatorcompare}

\begin{figure}[h!]
\centering
\includegraphics[width=0.48\textwidth]{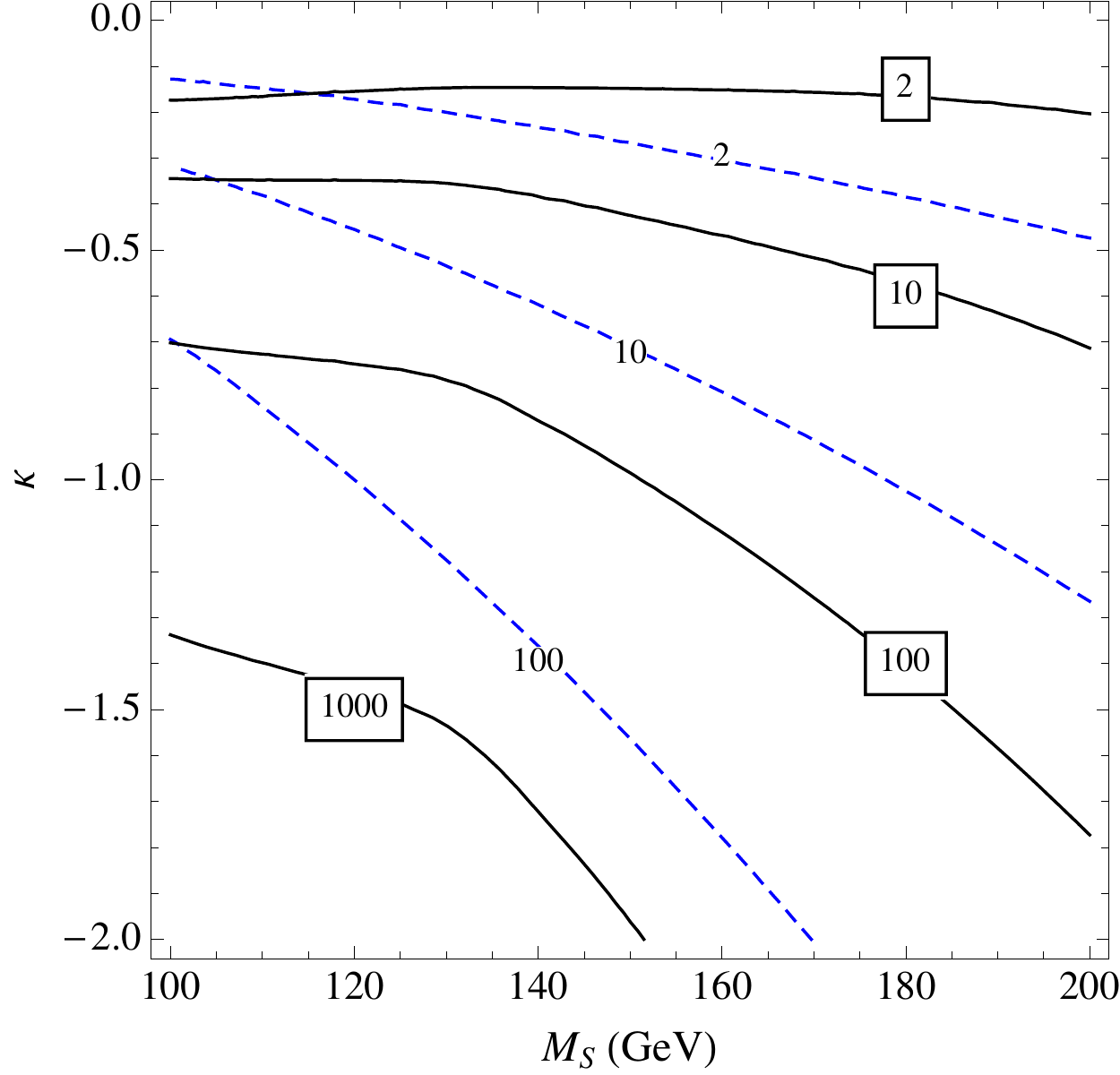}
\caption{Comparison of exact one-loop result against that 
obtained from using the effective couplings}
\label{fig:operatorcompare}
\end{figure}

Comparison of exact one-loop result against that 
obtained from using the effective couplings in 
Fig.~\ref{fig:operatorcompare}.
Here we only show a ``zoomed in'' region for small $M_S$
and negative $\kappa$, where the differences between
the calculations are largest.  Nevertheless, we have verified 
that for $M_i \gg v,\mh$, the contours asymptotically agree.

%\begin{figure*}
%\centering
%\includegraphics[width=3.0in]{mh125_scalvar.pdf} 
%\hspace{0.5in}
%\includegraphics[width=3.0in]{mh200_scalvar.pdf} \\
%\includegraphics[width=3.0in]{cont125_scalvar.pdf} 
%\hspace{0.5in}
%\includegraphics[width=3.0in]{cont200_scalvar.pdf}
%\caption{Factorization and renormalization scale dependence in the 
%di-Higgs rate for $\mh = 125\,\gev$. The ratio of the SM plus scalar 
%to SM alone $gg\ra hh$ cross section is shown for two different 
%scale choices: $\mu_F = \mu_R = 2\times \mh$ (solid) and 
%$\mu_F = \mu_R = \sqrt{\hat s}$ (dashed). 
%Going from $2\times \mh$ to $\sqrt{\hat s}$, the SM (only) 
%cross section changes by $O(10\%)$}
%\label{fig:scalvar}
%\end{figure*}

%\begin{figure}
%\centering
%\includegraphics[width=3.0in]{mh125_big.pdf}
%\caption{combined negative + positive $\kappa$ plot. Looks different 
%than the other $\kappa > 0$ plot because we plot $<\#$ in contours here, 
%rather than $>\#$.}
%\label{fig:bigmh125}
%\end{figure}

%------------------------------------------------------------------------------
\section{Renormalization Group Equations}
\label{sec:rgimp}

The renormalization group equations for $\kappa$, $\omega$, and $\lambda_h$
(taking $\mu_S$ to be small and negligible) are:
\begin{eqnarray}
16\pi^2 \frac{d\kappa}{dt} &=& 
    2\,\kappa^2 - 9 \kappa \, g_c^2 
    + 6 \, \kappa \, \lambda_h + 10 \, \kappa\, \omega \label{eq:kapparge} \\
16\pi^2 \frac{d\omega}{dt} &=& 
    \kappa^2 + 24 \, \omega^2 -18 \, g_c^2 \, \omega + 12 \, g_c^4
\end{eqnarray}
where $g_c$ is the $SU(3)$ coupling and $\lambda_h$ is the 
Standard Model Higgs quartic coupling.  All terms in the $\kappa$
renormalization group equation are proportional to $\kappa$, 
since $\kappa = 0$ is technically natural. 
The running of $\lambda_h$ is also changed by the Higgs 
portal interaction:
\begin{equation}
16\pi^2\frac{d\lambda_h}{dt} = \beta_{\lambda_h} + 4\,\kappa^2,
\end{equation}
where $\beta_{\lambda_h}$ is the $\beta$-function for the 
Higgs quartic coupling coming from SM interactions.

For large positive $\kappa$, Eq.~(\ref{eq:kapparge}) shows that
$\kappa$ will grow and ultimately hit a Landau pole, 
similar to what happens with the Higgs quartic coupling.  
For large negative $\kappa$, however, Eq.~(\ref{eq:kapparge}) 
shows that the Higgs-portal interaction rapidly \emph{decreases}
until the other terms in the renormalization group equation 
become important.  Hence, a one-loop renormalization group 
improved potential is necessary to analyze the stability 
of the effective potential in the presence of a large negative 
$\kappa$ coupling.  Our estimates suggest $\kappa \gsim -2$ 
is perfectly acceptable, however a detailed analysis of the 
scalar potential is beyond the scope of this paper.

%------------------------------------------------------------------------------
\section*{Acknowledgments}

We thank B.~Dobrescu for collaboration during the early part of this work,
and Y.~Gershtein for valuable conversations. 
GDK thanks the Theoretical Physics Group at Fermilab for 
warm hospitality where part of this work was completed.  
GDK was supported in part by the US Department of Energy under 
contract number DE-FG02-96ER40969 as well as a URA Visiting Scholar
Award from Fermilab. AM was supported by Fermilab 
operated by Fermi Research Alliance, LLC under contract 
number DE-AC02-07CH11359 with the US Department of Energy.

%------------------------------------------------------------------------------

%------------------------------------------------------------------------------


\begin{thebibliography}{99}

%\cite{Glover:1987nx}
\bibitem{Glover:1987nx}
  E.~W.~N.~Glover, J.~J.~van der Bij,
  %``Higgs Boson Pair Production Via Gluon Fusion,''
  Nucl.\ Phys.\  {\bf B309}, 282 (1988).
  
%\cite{Plehn:1996wb}
\bibitem{Plehn:1996wb} 
  T.~Plehn, M.~Spira and P.~M.~Zerwas,
  %``Pair production of neutral Higgs particles in gluon-gluon collisions,''
  Nucl.\ Phys.\ B {\bf 479}, 46 (1996)
  [Erratum-ibid.\ B {\bf 531}, 655 (1998)]
  [hep-ph/9603205].
  %%CITATION = HEP-PH/9603205;%%

%\cite{Dawson:1998py}
\bibitem{Dawson:1998py} 
  S.~Dawson, S.~Dittmaier and M.~Spira,
  %``Neutral Higgs boson pair production at hadron colliders: QCD corrections,''
  Phys.\ Rev.\ D {\bf 58}, 115012 (1998)
  [hep-ph/9805244].
  %%CITATION = HEP-PH/9805244;%%

%\cite{Djouadi:1999rca}
\bibitem{Djouadi:1999rca} 
  A.~Djouadi, W.~Kilian, M.~Muhlleitner and P.~M.~Zerwas,
  %``Production of neutral Higgs boson pairs at LHC,''
  Eur.\ Phys.\ J.\ C {\bf 10}, 45 (1999)
  [hep-ph/9904287].
  %%CITATION = HEP-PH/9904287;%%




%\cite{Barger:1988kb}
\bibitem{susysuppression}
%\cite{hep-ph/9508265}
%\bibitem{hep-ph/9508265} 
  G.~L.~Kane, G.~D.~Kribs, S.~P.~Martin and J.~D.~Wells,
  %``Two photon decays of the lightest Higgs boson of supersymmetry at the LHC,''
  Phys.\ Rev.\ D\ {\bf 53}, 213  (1996)
  [hep-ph/9508265]; 
  %%CITATION = PHRVA,D53,213;%%
%\cite{Dawson:1996xz}
%\bibitem{Dawson:1996xz} 
  S.~Dawson, A.~Djouadi and M.~Spira,
  %``QCD corrections to SUSY Higgs production: The Role of squark loops,''
  Phys.\ Rev.\ Lett.\  {\bf 77}, 16 (1996)
  [hep-ph/9603423].
  %%CITATION = HEP-PH/9603423;%%
%\cite{hep-ph/9806315}
%\bibitem{hep-ph/9806315} 
  A.~Djouadi,
  %``Squark effects on Higgs boson production and decay at the LHC,''
  Phys.\ Lett.\ B\ {\bf 435}, 101  (1998)
  [hep-ph/9806315];
  %%CITATION = PHLTA,B435,101;%%
    %\cite{hep-ph/0202167}
%\bibitem{hep-ph/0202167} 
  M.~S.~Carena, S.~Heinemeyer, C.~E.~M.~Wagner and G.~Weiglein,
 % ``Suggestions for benchmark scenarios for MSSM Higgs boson searches at hadron colliders,''
  Eur.\ Phys.\ J.\ C\ {\bf 26}, 601  (2003)
  [hep-ph/0202167];
    %%CITATION = EPHJA,C26,601;%%
%\cite{Muhlleitner:2006wx}
%\bibitem{Muhlleitner:2006wx} 
  M.~Muhlleitner and M.~Spira,
  %``Higgs Boson Production via Gluon Fusion: Squark Loops at NLO QCD,''
  Nucl.\ Phys.\ B {\bf 790}, 1 (2008)
  [hep-ph/0612254];
  %%CITATION = HEP-PH/0612254;%%
%\cite{arXiv:0901.0266}
%\bibitem{arXiv:0901.0266} 
  I.~Low and S.~Shalgar,
  %``Implications of the Higgs Discovery in the MSSM Golden Region,''
  JHEP\ {\bf 0904}, 091  (2009)
  [arXiv:0901.0266];
  %%CITATION = JHEPA,0904,091;%%
%\cite{Menon:2009mz}
%\bibitem{Menon:2009mz} 
  A.~Menon and D.~E.~Morrissey,
  %``Higgs Boson Signatures of MSSM Electroweak Baryogenesis,''
  Phys.\ Rev.\ D {\bf 79}, 115020 (2009)
  [arXiv:0903.3038];
  %%CITATION = ARXIV:0903.3038;%%

%\cite{Manohar:2006ga}
\bibitem{Manohar:2006ga}
  A.~V.~Manohar, M.~B.~Wise,
  %``Flavor changing neutral currents, an extended scalar sector, and the Higgs production rate at the LHC,''
  Phys.\ Rev.\  {\bf D74}, 035009 (2006)
  [hep-ph/0606172];
%\cite{He:2011ti}\bibitem{He:2011ti}
  X.~-G.~He, G.~Valencia,
  %``An extended scalar sector to address the tension between a fourth generation and Higgs searches at the LHC,''
  [arXiv:1108.0222].

%\cite{Arnesen:2008fb}
\bibitem{Arnesen:2008fb} 
  C.~Arnesen, I.~Z.~Rothstein and J.~Zupan,
  %``Smoking Guns for On-Shell New Physics at the LHC,''
  Phys.\ Rev.\ Lett.\  {\bf 103}, 151801 (2009)
  [arXiv:0809.1429].
  %%CITATION = ARXIV:0809.1429;%%

%\cite{Ma:2011kc}
\bibitem{Ma:2011kc} 
  E.~Ma, 
  %``Hiding the Higgs Boson from Prying Eyes,''
  Phys.\ Lett.\ B {\bf 706}, 350 (2012)
  [arXiv:1109.4177];
  %%CITATION = ARXIV:1109.4177;%%
%\cite{Ma:2011rp}\bibitem{Ma:2011rp}   E.~Ma,
  `%`Supersymmetric Axion-Neutrino Model with a Higgs Hybrid,''
  arXiv:1112.1367].
  %%CITATION = ARXIV:1112.1367;%%  

%\cite{arXiv:0709.4227}
\bibitem{arXiv:0709.4227} 
  R.~Bonciani, G.~Degrassi and A.~Vicini,
  %``Scalar particle contribution to Higgs production via gluon fusion at NLO,''
  JHEP\ {\bf 0711}, 095  (2007)
  [arXiv:0709.4227].
  %%CITATION = JHEPA,0711,095;%%
%\cite{Aglietti:2006tp}\bibitem{Aglietti:2006tp} 
  U.~Aglietti, R.~Bonciani, G.~Degrassi and A.~Vicini,
  %``Analytic Results for Virtual QCD Corrections to Higgs Production and Decay,''
  JHEP {\bf 0701}, 021 (2007)
  [hep-ph/0611266];
  %%CITATION = HEP-PH/0611266;%%
%\cite{Anastasiou:2006hc}\bibitem{Anastasiou:2006hc} 
  C.~Anastasiou, {\it et al}, %S.~Beerli, S.~Bucherer, A.~Daleo and Z.~Kunszt,
  %``Two-loop amplitudes and master integrals for the production of a Higgs boson via a massive quark and a scalar-quark loop,''
  JHEP {\bf 0701}, 082 (2007)
  [hep-ph/0611236].
  %%CITATION = HEP-PH/0611236;%%

%\cite{Boughezal:2010ry}
\bibitem{Boughezal:2010ry}
  R.~Boughezal, F.~Petriello,
  %``Color-octet scalar effects on Higgs boson production in gluon fusion,''
  Phys.\ Rev.\  {\bf D81}, 114033 (2010)
  [arXiv:1003.2046];
%\cite{arXiv:1101.3769}\bibitem{arXiv:1101.3769} 
  R.~Boughezal,
  %``Constraints on heavy colored scalars from Tevatron's Higgs exclusion limit,''
  Phys.\ Rev.\ D\ {\bf 83}, 093003  (2011)
  [arXiv:1101.3769].
  %%CITATION = PHRVA,D83,093003;%%

%\cite{Bai:2011aa}
\bibitem{Bai:2011aa} 
  Y.~Bai, J.~Fan and J.~L.~Hewett,
  %``Hiding a Heavy Higgs Boson at the 7 TeV LHC,''
  arXiv:1112.1964 [hep-ph].
  %%CITATION = ARXIV:1112.1964;%%

%\cite{Dobrescu:2011aa}
\bibitem{Dobrescu:2011aa} 
  B.~A.~Dobrescu, G.~D.~Kribs and A.~Martin,
  %``Higgs Underproduction at the LHC,''
  arXiv:1112.2208 [hep-ph].
  %%CITATION = ARXIV:1112.2208;%%

%\cite{Batell:2011pz}
\bibitem{Batell:2011pz} 
  B.~Batell, S.~Gori and L.~-T.~Wang,
  %``Exploring the Higgs Portal with 10/fb at the LHC,''
  JHEP {\bf 1206}, 172 (2012)
  [arXiv:1112.5180 [hep-ph]].
  %%CITATION = ARXIV:1112.5180;%%

%\cite{Aad:2011yh}
\bibitem{Barger:1988kb} 
  V.~D.~Barger and T.~Han,
  %``DOUBLE HIGGS BOSON PRODUCTION VIA W W FUSION IN TeV e+ e- COLLISIONS,''
  Mod.\ Phys.\ Lett.\ A {\bf 5}, 667 (1990).
  %%CITATION = MPLAE,A5,667;%%

%\cite{Hagiwara:1989xx}
\bibitem{Hagiwara:1989xx} 
  K.~Hagiwara and H.~Murayama,
  %``Multiple Weak Boson Production Via Gluon Fusion,''
  Phys.\ Rev.\ D {\bf 41}, 1001 (1990).
  %%CITATION = PHRVA,D41,1001;%%

%\cite{Barger:1991jn}
\bibitem{Barger:1991jn} 
  V.~D.~Barger, A.~L.~Stange and R.~J.~N.~Phillips,
  %``Multiple weak boson signals at hadron supercolliders,''
  Phys.\ Rev.\ D {\bf 45}, 1484 (1992).
  %%CITATION = PHRVA,D45,1484;%%

%\cite{Djouadi:1999gv}
\bibitem{Djouadi:1999gv} 
  A.~Djouadi, W.~Kilian, M.~Muhlleitner and P.~M.~Zerwas,
  %``Testing Higgs selfcouplings at e+ e- linear colliders,''
  Eur.\ Phys.\ J.\ C {\bf 10}, 27 (1999)
  [hep-ph/9903229].
  %%CITATION = HEP-PH/9903229;%%

%\cite{Belyaev:1999mx}
\bibitem{Belyaev:1999mx} 
  A.~Belyaev, M.~Drees, O.~J.~P.~Eboli, J.~K.~Mizukoshi and S.~F.~Novaes,
  %``Supersymmetric Higgs pair production at hadron colliders,''
  Phys.\ Rev.\ D {\bf 60}, 075008 (1999)
  [hep-ph/9905266].
  %%CITATION = HEP-PH/9905266;%%

%\cite{BarrientosBendezu:2001di}
\bibitem{BarrientosBendezu:2001di} 
  A.~A.~Barrientos Bendezu and B.~A.~Kniehl,
  %``Pair production of neutral Higgs bosons at the CERN large hadron collider,''
  Phys.\ Rev.\ D {\bf 64}, 035006 (2001)
  [hep-ph/0103018].
  %%CITATION = HEP-PH/0103018;%%

%\cite{Baur:2002rb}
\bibitem{Baur:2002rb} 
  U.~Baur, T.~Plehn and D.~L.~Rainwater,
  %``Measuring the Higgs boson self coupling at the LHC and finite top mass matrix elements,''
  Phys.\ Rev.\ Lett.\  {\bf 89}, 151801 (2002)
  [hep-ph/0206024].
  %%CITATION = HEP-PH/0206024;%%

%\cite{Baur:2002qd}
\bibitem{Baur:2002qd} 
  U.~Baur, T.~Plehn and D.~L.~Rainwater,
  %``Determining the Higgs boson selfcoupling at hadron colliders,''
  Phys.\ Rev.\ D {\bf 67}, 033003 (2003)
  [hep-ph/0211224].
  %%CITATION = HEP-PH/0211224;%%

%\cite{Baur:2003gp}
\bibitem{Baur:2003gp} 
  U.~Baur, T.~Plehn and D.~L.~Rainwater,
  %``Probing the Higgs selfcoupling at hadron colliders using rare decays,''
  Phys.\ Rev.\ D {\bf 69}, 053004 (2004)
  [hep-ph/0310056].
  %%CITATION = HEP-PH/0310056;%%

%\cite{Moretti:2004wa}
\bibitem{Moretti:2004wa} 
  M.~Moretti, S.~Moretti, F.~Piccinini, R.~Pittau and A.~D.~Polosa,
  %``Higgs boson self-couplings at the LHC as a probe of extended Higgs sectors,''
  JHEP {\bf 0502}, 024 (2005)
  [hep-ph/0410334].
  %%CITATION = HEP-PH/0410334;%%

%\cite{Plehn:2005nk}
\bibitem{Plehn:2005nk} 
  T.~Plehn and M.~Rauch,
  %``The quartic higgs coupling at hadron colliders,''
  Phys.\ Rev.\ D {\bf 72}, 053008 (2005)
  [hep-ph/0507321].
  %%CITATION = HEP-PH/0507321;%%

%\cite{Binoth:2006ym}
\bibitem{Binoth:2006ym} 
  T.~Binoth, S.~Karg, N.~Kauer and R.~Ruckl,
  %``Multi-Higgs boson production in the Standard Model and beyond,''
  Phys.\ Rev.\ D {\bf 74}, 113008 (2006)
  [hep-ph/0608057].
  %%CITATION = HEP-PH/0608057;%%

%\cite{Pierce:2006dh}
\bibitem{Pierce:2006dh}
  A.~Pierce, J.~Thaler, L.~-T.~Wang,
  %``Disentangling dimension six operators through di-Higgs boson production,''
  JHEP {\bf 0705}, 070 (2007)
  [hep-ph/0609049].

%\cite{Dawson:2006dm}
\bibitem{Dawson:2006dm} 
  S.~Dawson, C.~Kao, Y.~Wang and P.~Williams,
  %``QCD Corrections to Higgs Pair Production in Bottom Quark Fusion,''
  Phys.\ Rev.\ D {\bf 75}, 013007 (2007)
  [hep-ph/0610284].
  %%CITATION = HEP-PH/0610284;%%

%\cite{Moretti:2007ca}
\bibitem{Moretti:2007ca} 
  M.~Moretti, S.~Moretti, F.~Piccinini, R.~Pittau and J.~Rathsman,
  %``Vector-Boson Production of Light Higgs Pairs in 2-Higgs Doublet Models,''
  JHEP {\bf 0712}, 075 (2007)
  [arXiv:0706.4117 [hep-ph]].
  %%CITATION = ARXIV:0706.4117;%%
  
%\cite{Kanemura:2008ub}
\bibitem{Kanemura:2008ub}
  S.~Kanemura, K.~Tsumura,
  %``Effects of the anomalous Higgs couplings on the Higgs boson production at the LHC,''
  Eur.\ Phys.\ J.\  {\bf C63}, 11-21 (2009)
  [arXiv:0810.0433].

%\cite{Lafaye:2009vr}
\bibitem{Lafaye:2009vr} 
  R.~Lafaye, T.~Plehn, M.~Rauch, D.~Zerwas and M.~Duhrssen,
  %``Measuring the Higgs Sector,''
  JHEP {\bf 0908}, 009 (2009)
  [arXiv:0904.3866 [hep-ph]].
  %%CITATION = ARXIV:0904.3866;%%

%\cite{Arhrib:2009hc}
\bibitem{Arhrib:2009hc} 
  A.~Arhrib, R.~Benbrik, C.~-H.~Chen, R.~Guedes and R.~Santos,
  %``Double Neutral Higgs production in the Two-Higgs doublet model at the LHC,''
  JHEP {\bf 0908}, 035 (2009)
  [arXiv:0906.0387 [hep-ph]].
  %%CITATION = ARXIV:0906.0387;%%

%\cite{Moretti:2010kc}
\bibitem{Moretti:2010kc} 
  M.~Moretti, S.~Moretti, F.~Piccinini, R.~Pittau and J.~Rathsman,
  %``Production of Light Higgs Pairs in 2-Higgs Doublet Models via the Higgs-strahlung Process at the LHC,''
  JHEP {\bf 1011}, 097 (2010)
  [arXiv:1008.0820 [hep-ph]].
  %%CITATION = ARXIV:1008.0820;%%

%\cite{Asakawa:2010xj}
\bibitem{Asakawa:2010xj} 
  E.~Asakawa, D.~Harada, S.~Kanemura, Y.~Okada and K.~Tsumura,
  %``Higgs boson pair production in new physics models at hadron, lepton, and photon colliders,''
  Phys.\ Rev.\ D {\bf 82}, 115002 (2010)
  [arXiv:1009.4670 [hep-ph]].
  %%CITATION = ARXIV:1009.4670;%%

%\cite{Grober:2010yv}
\bibitem{Grober:2010yv} 
  R.~Grober and M.~Muhlleitner,
  %``Composite Higgs Boson Pair Production at the LHC,''
  JHEP {\bf 1106}, 020 (2011)
  [arXiv:1012.1562 [hep-ph]].
  %%CITATION = ARXIV:1012.1562;%%

%\cite{Contino:2012xk}
\bibitem{Contino:2012xk} 
  R.~Contino, M.~Ghezzi, M.~Moretti, G.~Panico, F.~Piccinini and A.~Wulzer,
  %``Anomalous Couplings in Double Higgs Production,''
  arXiv:1205.5444 [hep-ph].
  %%CITATION = ARXIV:1205.5444;%%

%\cite{Dolan:2012rv}
\bibitem{Dolan:2012rv} 
  M.~J.~Dolan, C.~Englert and M.~Spannowsky,
  %``Higgs self-coupling measurements at the LHC,''
  arXiv:1206.5001 [hep-ph].
  %%CITATION = ARXIV:1206.5001;%%
















\bibitem{Aad:2011yh}
  G.~Aad {\it et al.} [ATLAS Collaboration],
  %``Search for massive colored scalars in four-jet final states in 
    %$\sqrt{s}=7$~TeV proton-proton collisions,''
  [arXiv:1110.2693].
  
\bibitem{CMSoctets}
CMS Collaboration,
CMS PAS EXO-11-016
 











%\cite{Bai:2010dj}
\bibitem{atlashiggs}
  G.~Aad {\it et al.}  [ATLAS Collaboration],
  %``Combined search for the Standard Model Higgs boson in pp collisions at sqrt(s) = 7 TeV with the ATLAS detector,''
  arXiv:1207.0319 [hep-ex].
  %%CITATION = ARXIV:1207.0319;%%

%\cite{cmshiggs}
\bibitem{cmshiggs}
  CMS Collaboration,
  %''Observation of a new boson with a mass near 125 GeV''
  CMS-PAS-HIG-12-020.






%\cite{Ellis:1975ap}
\bibitem{Higgsimpostors}
  W.~D.~Goldberger and M.~B.~Wise,
  %``Phenomenology of a stabilized modulus,''
  Phys.\ Lett.\  B {\bf 475}, 275 (2000)
  [arXiv:hep-ph/9911457],
  %%CITATION = PHLTA,B475,275;%%
%\cite{Giudice:2000av}
%\bibitem{Giudice:2000av}
  G.~F.~Giudice, R.~Rattazzi and J.~D.~Wells,
  %``Graviscalars from higher dimensional metrics and curvature Higgs mixing,''
  Nucl.\ Phys.\  B {\bf 595}, 250 (2001)
  [arXiv:hep-ph/0002178],
  %%CITATION = NUPHA,B595,250;%%
  %\cite{Csaki:2000zn}
%\bibitem{Csaki:2000zn}
  C.~Csaki, M.~L.~Graesser and G.~D.~Kribs,
  %``Radion dynamics and electroweak physics,''
  Phys.\ Rev.\  D {\bf 63}, 065002 (2001)
  [arXiv:hep-th/0008151],
  %%CITATION = PHRVA,D63,065002;%%
  %\cite{Goldberger:2007zk}
%\bibitem{Goldberger:2007zk}
  W.~D.~Goldberger, B.~Grinstein and W.~Skiba,
  %``Distinguishing the Higgs boson from the dilaton at the Large Hadron
  %Collider,''
  Phys.\ Rev.\ Lett.\  {\bf 100} (2008) 111802
  [arXiv:0708.1463 [hep-ph]],
  %%CITATION = PRLTA,100,111802;%%
%\cite{DeRujula:2010ys}
%\bibitem{DeRujula:2010ys}
  A.~De Rujula, J.~Lykken, M.~Pierini, C.~Rogan and M.~Spiropulu,
  %``Higgs look-alikes at the LHC,''
  Phys.\ Rev.\  D {\bf 82}, 013003 (2010)
  [arXiv:1001.5300 [hep-ph]],
  %%CITATION = PHRVA,D82,013003;%%
%\cite{Low:2010jp}
%\bibitem{Low:2010jp}
  I.~Low and J.~Lykken,
  %``Revealing the electroweak properties of a new scalar resonance,''
  JHEP {\bf 1010}, 053 (2010)
  [arXiv:1005.0872 [hep-ph]],
  %%CITATION = JHEPA,1010,053;%%
%\cite{Davoudiasl:2010fb}
%\bibitem{Davoudiasl:2010fb}
  H.~Davoudiasl, T.~McElmurry and A.~Soni,
  %``Promising Diphoton Signals of the Little Radion at Hadron Colliders,''
  Phys.\ Rev.\  D {\bf 82}, 115028 (2010)
  [arXiv:1009.0764 [hep-ph]],
  %%CITATION = PHRVA,D82,115028;%%
%\cite{Fox:2011qc}
%\bibitem{Fox:2011qc}
  P.~J.~Fox, D.~Tucker-Smith and N.~Weiner,
  %``Higgs friends and counterfeits at hadron colliders,''
  JHEP {\bf 1106}, 127 (2011)
  [arXiv:1104.5450 [hep-ph]].
  %%CITATION = JHEPA,1106,127;%%

%\cite{Hagiwara:1990dw}
\bibitem{Ellis:1975ap} 
  J.~R.~Ellis, M.~K.~Gaillard and D.~V.~Nanopoulos,
  %``A Phenomenological Profile of the Higgs Boson,''
  Nucl.\ Phys.\ B {\bf 106}, 292 (1976).
  %%CITATION = NUPHA,B106,292;%%

%\cite{Shifman:1979eb}
\bibitem{Shifman:1979eb} 
  M.~A.~Shifman, A.~I.~Vainshtein, M.~B.~Voloshin and V.~I.~Zakharov,
  %``Low-Energy Theorems for Higgs Boson Couplings to Photons,''
  Sov.\ J.\ Nucl.\ Phys.\  {\bf 30}, 711 (1979)
  [Yad.\ Fiz.\  {\bf 30}, 1368 (1979)].
  %%CITATION = SJNCA,30,711;%%

%\cite{Kniehl:1995tn}
\bibitem{Kniehl:1995tn} 
  B.~A.~Kniehl and M.~Spira,
  %``Low-energy theorems in Higgs physics,''
  Z.\ Phys.\ C {\bf 69}, 77 (1995)
  [hep-ph/9505225].
  %%CITATION = HEP-PH/9505225;%%

%\cite{Low:2009di}
\bibitem{Low:2009di} 
  I.~Low, R.~Rattazzi and A.~Vichi,
  %``Theoretical Constraints on the Higgs Effective Couplings,''
  JHEP {\bf 1004}, 126 (2010)
  [arXiv:0907.5413 [hep-ph]].
  %%CITATION = ARXIV:0907.5413;%%

%\cite{Gillioz:2012se}
\bibitem{Gillioz:2012se} 
  M.~Gillioz, R.~Grober, C.~Grojean, M.~Muhlleitner and E.~Salvioni,
  %``Higgs Low-Energy Theorem (and its corrections) in Composite Models,''
  arXiv:1206.7120 [hep-ph].
  %%CITATION = ARXIV:1206.7120;%%


%\cite{Butterworth:2008iy}
\bibitem{Bai:2010dj}
  Y.~Bai, B.~A.~Dobrescu,
 % ``Heavy octets and Tevatron signals with three or four b jets,''
  JHEP {\bf 1107}, 100 (2011).
  [arXiv:1012.5814].

%\cite{HUTP-91-A009}
\bibitem{HUTP-91-A009}
  R.~S.~Chivukula, M.~Golden, E.~H.~Simmons,
  %``Multi-jet physics at hadron colliders'',
  Nucl. Phys. B{\bf 363} (1991) $\!$83.
  %%CITATION = NUPHA,B363,83;%%

%\cite{Dobrescu:2007yp}
\bibitem{Dobrescu:2007yp}
  B.~A.~Dobrescu, K.~Kong, R.~Mahbubani,
  %``Massive color-octet bosons and pairs of resonances at hadron colliders,''
  Phys.\ Lett.\  {\bf B670}, 119-123 (2008)
  [arXiv:0709.2378];
%\cite{Arnold:2011ra}\bibitem{Arnold:2011ra} 
  J.~M.~Arnold and B.~Fornal,
  %``Color octet scalars and high pT four-jet events at LHC,''
  arXiv:1112.0003.
  %%CITATION = ARXIV:1112.0003;%%

%\cite{Mantry:2007ar}
\bibitem{Alwall:2007st} 
  J.~Alwall, P.~Demin, S.~de Visscher, R.~Frederix, M.~Herquet, F.~Maltoni, T.~Plehn and D.~L.~Rainwater {\it et al.},
  %``MadGraph/MadEvent v4: The New Web Generation,''
  JHEP {\bf 0709}, 028 (2007)
  [arXiv:0706.2334 [hep-ph]].
  %%CITATION = ARXIV:0706.2334;%%

%\cite{atlashiggs}
\bibitem{Hagiwara:1990dw} 
  K.~Hagiwara, H.~Murayama and I.~Watanabe,
  %``Search for the Yukawa interaction in the process e+ e- ---> t anti-t Z at TeV linear colliders,''
  Nucl.\ Phys.\ B {\bf 367}, 257 (1991).
  %%CITATION = NUPHA,B367,257;%%
  
  %\cite{Murayama:1992gi}
\bibitem{Murayama:1992gi} 
  H.~Murayama, I.~Watanabe and K.~Hagiwara,
  %``HELAS: HELicity amplitude subroutines for Feynman diagram evaluations,''
  KEK-91-11.
  %%CITATION = KEK-91-11;%%


\bibitem{Hahn:1998yk} 
  T.~Hahn and M.~Perez-Victoria,
  ``Automatized one loop calculations in four-dimensions and D-dimensions,''
  Comput.\ Phys.\ Commun.\  {\bf 118}, 153 (1999)
  [hep-ph/9807565].
  %%CITATION = HEP-PH/9807565;%%
  
  %\cite{Alwall:2007st}
\bibitem{Butterworth:2008iy} 
  J.~M.~Butterworth, A.~R.~Davison, M.~Rubin and G.~P.~Salam,
  %``Jet substructure as a new Higgs search channel at the LHC,''
  Phys.\ Rev.\ Lett.\  {\bf 100}, 242001 (2008)
  [arXiv:0802.2470 [hep-ph]].
  %%CITATION = ARXIV:0802.2470;%%

%\cite{Abdesselam:2010pt}
\bibitem{Abdesselam:2010pt} 
  A.~Abdesselam, E.~B.~Kuutmann, U.~Bitenc, G.~Brooijmans, J.~Butterworth, P.~Bruckman de Renstrom, D.~Buarque Franzosi and R.~Buckingham {\it et al.},
  %``Boosted objects: A Probe of beyond the Standard Model physics,''
  Eur.\ Phys.\ J.\ C {\bf 71}, 1661 (2011)
  [arXiv:1012.5412 [hep-ph]].
  %%CITATION = ARXIV:1012.5412;%%

%\cite{Altheimer:2012mn}
\bibitem{Altheimer:2012mn} 
  A.~Altheimer, S.~Arora, L.~Asquith, G.~Brooijmans, J.~Butterworth, M.~Campanelli, B.~Chapleau and A.~E.~Cholakian {\it et al.},
  %``Jet Substructure at the Tevatron and LHC: New results, new tools, new benchmarks,''
  J.\ Phys.\ G G {\bf 39}, 063001 (2012)
  [arXiv:1201.0008 [hep-ph]].
  %%CITATION = ARXIV:1201.0008;%%

%\cite{Katz:2010iq}
\bibitem{Katz:2010iq} 
  A.~Katz, M.~Son and B.~Tweedie,
  %``Ditau-Jet Tagging and Boosted Higgses from a Multi-TeV Resonance,''
  Phys.\ Rev.\ D {\bf 83}, 114033 (2011)
  [arXiv:1011.4523 [hep-ph]].
  %%CITATION = ARXIV:1011.4523;%%
  
  
  %\cite{Englert:2011iz}
\bibitem{Englert:2011iz}
  C.~Englert, T.~S.~Roy and M.~Spannowsky,
  %``Ditau jets in Higgs searches,''
  Phys.\ Rev.\ D {\bf 84} (2011) 075026
  [arXiv:1106.4545 [hep-ph]].
  %%CITATION = ARXIV:1106.4545;%%

\bibitem{Mantry:2007ar} 
  S.~Mantry, M.~Trott and M.~B.~Wise,
  %``The Higgs decay width in multi-scalar doublet models,''
  Phys.\ Rev.\ D {\bf 77}, 013006 (2008)
  [arXiv:0709.1505 [hep-ph]].
  %%CITATION = ARXIV:0709.1505;%%
  
  %\cite{Arnold:2009ay}
\bibitem{Arnold:2009ay} 
  J.~M.~Arnold, M.~Pospelov, M.~Trott and M.~B.~Wise,
  %``Scalar Representations and Minimal Flavor Violation,''
  JHEP {\bf 1001}, 073 (2010)
  [arXiv:0911.2225 [hep-ph]].
  %%CITATION = ARXIV:0911.2225;%%
  

\end{thebibliography}
\end{document}